\documentstyle[epsfig]{mn}
\newif\ifAMStwofonts
\AMStwofontstrue                

\def\lm{_{\ell m}}

\DeclareMathVersion{bold}                    
\DeclareMathAlphabet{\mathbfit}{OT1}{cmr}{bx}{it}
\SetMathAlphabet\mathbfit{bold}{OT1}{cmr}{bx}{it}
\DeclareMathAlphabet{\mathbfss}{OT1}{cmss}{bx}{n}
\SetMathAlphabet\mathbfss{bold}{OT1}{cmss}{bx}{n}
\ifAMStwofonts
  \ifCUPmtlplainloaded \else
    \DeclareSymbolFont{UPM}{U}{eur}{m}{n}
    \SetSymbolFont{UPM}{bold}{U}{eur}{b}{n}
    \DeclareSymbolFont{AMSa}{U}{msa}{m}{n}
    \DeclareMathSymbol{\upi}{0}{UPM}{"19}
    \DeclareMathSymbol{\umu}{0}{UPM}{"16}
    \DeclareMathSymbol{\upartial}{0}{UPM}{"40}
    \DeclareMathSymbol{\leqslant}{3}{AMSa}{"36}
    \DeclareMathSymbol{\geqslant}{3}{AMSa}{"3E}

     \let\le=\leqslant

  \fi
\fi

\title[Flexible MEM]
{Foreground separation using a flexible maximum-entropy algorithm: an
application to COBE data}
\author[Barreiro et al.]{R.B.~Barreiro$^1$, M.P.~Hobson$^2$, A.J.~Banday$^3$,
A.N.~Lasenby$^2$, V.~Stolyarov$^2$, 
\newauthor P.~Vielva$^1$ and K.M.~G\'orski$^4$ \\
$^1$Instituto de F\'\i sica de Cantabria, CSIC-Universidad de Cantabria,
Santander, Spain \\
$^2$Astrophysics Group, Cavendish Laboratory, Madingley Road,
Cambridge CB3 0HE \\
$^3$Max-Planck Institut f\"ur Astrophysik (MPA), Karl-Schwarzschild
Strasse 1, D-85740 Garching, Germany \\
$^4$European Southern Observatory (ESO), Karl-Schwarzschild Strasse 2,
D-85740 Garching, Germany }
\date{Accepted ---. Received ---; in original form \today}
\pubyear{2003}

\begin{document}
\maketitle
\begin{abstract}
A flexible maximum-entropy component separation algorithm is presented
that accommodates anisotropic noise, incomplete
sky-coverage and uncertainties in the
spectral parameters of foregrounds. The capabilities of the method 
are determined by first applying it to simulated spherical microwave 
data sets emulating the COBE-DMR, COBE-DIRBE and Haslam surveys. 
Using these simulations we find that is very difficult to
determine unambiguously the spectral parameters of the galactic
components for this data set due to their high level of
noise. Nevertheless, we show that is possible to find a robust CMB
reconstruction, especially at the high galactic latitude. 
The method is then applied to these real data sets to obtain
reconstructions of the CMB component and galactic foreground emission
over the whole sky. The best reconstructions are found for values of
the spectral parameters: $T_d=19$ K, $\alpha_d=2$, $\beta_{\rm
ff}=-0.19$ and $\beta_{\rm syn}=-0.8$. The CMB map has been recovered
with an estimated statistical error of $\sim 22 \mu$K on an angular
scale of $7$ degrees outside the galactic cut whereas the low galactic
latitude region presents contamination from the 
foreground emissions.
\end{abstract}


\section{Introduction}
Observation of the cosmic microwave background (CMB) radiation is one
of the most powerful tools of modern cosmology.  A handful of experiments
-- Boomerang (Netterfield et al. 2002), MAXIMA (Hanany et al. 2000),
DASI (Halverson et al. 2002), VSA (Grainge et al. 2003), 
CBI (Mason et al. 2003), ACBAR (Kuo et al. 2004), Archeops (Benoit
et al. 2003), WMAP (Bennett et al. 2003a) -- have already reported the
measurement of CMB fluctuations at subdegree scales, allowing tight
constraints to be placed on cosmological parameters.  Moreover,
current and future CMB experiments will measure the fluctuations with
unprecedented resolution, sensitivity, sky and frequency
coverage. Most notably, these include the WMAP mission by NASA
(that will continue to take data in the next years) and the
Planck mission by ESA (to be launched
in 2007), both of which will provide all-sky multifrequency
observations of the CMB.

When measuring the microwave sky, however, one does not only receive the
cosmological signal but also emission from our own Galaxy,
thermal and kinetic Sunyaev-Zeldovich emissions from clusters of
galaxies and emission from extragalactic point sources. In addition,
instrumental noise and possibly some systematic effect will
contaminate the data. Therefore, our capacity to extract all the
valuable information encoded in the CMB anisotropies will depend 
critically on our ability to separate the cosmological signal from the 
other microwave components.

Different methods have been proposed in the literature to perform this
component separation. Some techniques attempt to separate and
reconstruct all the components at the same time, such as Wiener
filtering (Bouchet, Gispert \& Puget 1996, Tegmark \& Efstathiou
1996), the maximum-entropy method (Hobson et al. 1998, Stolyarov et al
2002) or blind source separation (Baccigalupi et al. 2000, Maino
et al. 2002, Delabrouille et al. 2003). 
Another approach is to extract only the component of
interest from the data, as is the case, for example, when one is
searching for emission from extragalactic point sources or the
Sunyaev-Zel'dovich effect in clusters. Methods for performing such a
separation include the use of the mexican hat wavelet (Cay\'on et
al. 2000, Vielva et al. 2001a, Vielva et al. 2003) matched
filters (e.g. Tegmark \& de Oliveira-Costa 1998),
scale-adaptive filters (Sanz, Herranz \&
Mart\'{\i}nez-Gonz\'alez 2001, Herranz et al. 2002a,b,c) -- see also
Barreiro et al. (2003) for a comparison of the performance of these
filters --,
the McClean algorithm (Hobson \& McLachlan 2003), the Bayesian
approach proposed by Diego et al. (2002) or the blind EM algorithm of
Mart\'\i nez-Gonz\'alez et al. (2003). 

In the present work, we will focus on the maximum-entropy method
(MEM) for reconstruction of all components simultaneously. This 
technique has been successfully applied in reconstructing
microwave components from simulated Planck data in a small patch of
the sky (Hobson et al. 1998, hereafter H98). This work was extended to
deal with point sources (Hobson et al. 1999) and also combined with the
mexican hat wavelet (MHW; Vielva et al. 2001b). Recently, the MEM and
MEM+MHW algorithms have been adapted to deal with spherical data at
full Planck resolution by Stolyarov et al. (2002) (hereafter S02) and
Stolyarov et al. (2003) respectively.  The algorithm is capable of
analysing the vast amount of data expected from the Planck mission by
working in harmonic space, assuming independence between different
harmonic coefficients. Although this assumption is not necessarily
true, the method is very successful in performing a full-sky component
separation. Nevertheless, it would be desirable to find a method that
can perform this task without the assumption of independent harmonic
modes, and can be applied directly in the space where the data have
been observed. This would also allow one to introduce the properties
of the noise in a more straightforward manner and to deal simply with
incomplete and arbitrary shape sky coverage.

Another shortcoming of the standard MEM approach to component
separation is that the spectral dependence of the microwave components
needs to be known a priori. Although this is the case for the CMB and
the kinetic and thermal SZ effects, the spectral behaviour of the
galactic foregrounds is only approximately known. Moreover, it varies
with position and frequency.  The point sources, in particular, can
cause problems, since each source will have a different spectral
behaviour. This last point has been solved by combining MEM with the
MHW, which is optimised for the detection of point sources.
Nevertheless, the standard approach still lacks a way to estimate the
spectral dependence of the diffuse components present in the data.

In this paper, we present a maximum-entropy component separation
method that works in both real and harmonic space and is able to deal
with many of the problems mentioned above. Our analysis also includes
a thorough study of the properties of the reconstructions to estimate
the (average) spectral parameters for the galactic components.
Unfortunately, the price one has to pay for this flexible MEM is to
make the method much slower than the harmonic MEM used in S02.  To
illustrate the performance of the algorithm, we apply it first to
simulated spherical data and then to a set of real spherical data
available during the development of the algorithm, including the
COBE-DMR, COBE-DIRBE and Haslam maps, from which we reconstruct the
CMB emission and galactic foreground components.

The outline of the paper is as follows. In Section~\ref{sec:method} we
outline the 
flexible maximum-entropy component separation method. The spherical
microwave data set that we have analysed is described in
Section~\ref{sec:data}. Sections~\ref{sec:results_simu} and
\ref{sec:results_data} are devoted to the analyses of simulated
(including an artificially low noise case) and
real data respectively. Finally, we present our discussion and
conclusions in Section~\ref{sec:discussion}.

\section{The separation algorithm}
\label{sec:method}

In this section we outline the basics of our separation method, 
focusing particularly on the differences between the 
algorithm used in this work and traditional harmonic-based MEM
component separation.  For a more detailed derivation of MEM, see 
Hobson et al. (1998) and Stolyarov et al. (2002).

\subsection{The problem}
Our aim is to reconstruct the CMB anisotropies and the foreground
components in the presence of instrumental noise from multifrequency
microwave observations. We will assume that point sources have been previously
subtracted using the MHW, or other filtering technique, and that their
residual contribution is negligible.
If we observe the microwave sky in a given direction $\mathbf{x}$ at $n_f$
frequencies, we obtain a $n_f$-dimensional data vector that contains the
observed temperature fluctuations in this direction at each observing
frequency, plus instrumental noise. The observed data at the $\nu$th
frequency in the direction $\mathbf{x}$ can be written as
\begin{equation}
d_{\nu}(\mathbf{x})= \sum_{j=1}^{N_p}B_\nu(|\mathbf{x}-\mathbf{x}_j|)
\sum_{p=1}^{n_c} F_{\nu p}\,s_p(\mathbf{x}_j)\,
+ \epsilon_\nu(\mathbf{x}),
\label{datadef}
\end{equation}
where $N_p$ denotes the number of pixels in each map and $n_c$ is the
number of physical components to be separated. As is usual for the
MEM algorithm, we make the assumption that each of the components can
be factorised into a spatial template ($s_p$) at a reference frequency
$\nu_0$ and a frequency dependence encoded in $F_{\nu p}$. The
function $B_\nu$ accounts for the instrumental beam and 
$\epsilon_{\nu}(\mathbf{x})$ corresponds to
the instrumental noise at frequency $\nu$ and position $\mathbf{x}$.

\subsection{Harmonic-space MEM}

If the beam is circularly symmetric and $d_{\nu}$ is measured over the
whole sky, it is convenient to write the former equation in harmonic
space as
\begin{equation}
d_{\ell m}^{\nu}=\sum_{p=1}^{n_c} R_{\ell}^{\nu p} a_{\ell m}^p 
+ \epsilon_{\ell m}^{\nu},
\end{equation}
\label{datalm}
where we have adopted the usual notation for spherical harmonic
coefficients $f_{\ell m}=\int_{4 \pi} d\Omega Y_{\ell m}^*({\mathbf x}) 
f({\mathbf x}) $ in which $Y_{\ell m}({\mathbf x})$ is a standard
spherical harmonic function. 
Therefore, $d_{\ell m}^{\nu}$ and $\epsilon_{\ell m}^{\nu}$
correspond to the spherical harmonic coefficients of the data and the
noise at the $\nu$th frequency respectively, whereas $a_{\ell m}^p$ are
the harmonic coefficients of the $p$ component at the reference
frequency. The response matrix $R_{\ell}^{\nu p}=B_{\ell} F_{\nu p}$
(where $B_{\ell}$ are the harmonic coefficients of the $\nu$th
observing beam) determines the contribution of each physical component
to the data. Using matrix notation, we have for each mode
\begin{equation}
{\mathbf d}\lm={\mathbf R}_{\ell}{\mathbf a}_{\ell m} + \bepsilon_{\ell m},
\end{equation}
where ${\mathbf d}_{\ell m}$, ${\mathbf a}_{\ell m}$ and 
$\bepsilon_{\ell m}$ are column vectors of dimension 
$n_f$, $n_c$ and $n_f$ complex components
respectively, whereas the response matrix ${\mathbf R}_{\ell}$ 
contains $n_f \times n_c$ elements. 

If one neglects any correlations between different $(\ell,m)$ modes, the
reconstruction can be performed mode-by-mode, which greatly simplifies the
problem. This corresponds to assuming that the emission from each 
physical component and the instrumental noise are isotropic random
fields on the sky. Although, in reality, this is not the case, 
and so correlations do exist between different modes, the mode-by-mode 
harmonic-based MEM produces
excellent reconstructions. In this approximation, the a priori 
covariance structure of the different components is assumed to take
the form
\begin{equation}
\langle {\mathbf a}_{\ell m} {\mathbf a}_{\ell'm'} \rangle ={\mathbf C}_\ell
\delta_{\ell \ell'} \delta_{m m'},
\end{equation}
where we have made the additional simplifying assumption that the
covariance matrix does depend only on $\ell$.
Analogously, the (cross) power spectra for the noise can be written as
\begin{equation}
\langle\bepsilon_{\ell m} \bepsilon_{\ell' m'}\rangle
={\mathbf N}_\ell \delta_{\ell \ell'} \delta_{m m'}
\end{equation}
Thus, although the algorithm assumes statistical isotropy, it
can straightforwardly include any a priori knowledge of the (cross)
power spectra of the physical components and the noise at different
observing frequencies.

As explained in H98, any a priori power spectrum information
can be provided to the MEM algorithm by introducing a `hidden' 
vector ${\mathbf h}_{\ell m}$ that is a priori
uncorrelated (see H98) and relates to the signal through
\begin{equation}
{\mathbf a}_{\ell m}={\mathbf L}_{\ell} {\mathbf h}_{\ell m},
\label{hidden}
\end{equation}
where the lower triangular matrix ${\mathbf L}_{\ell}$ is obtained by
performing the Cholesky decomposition of the cross power spectra
${\mathbf C}_\ell$ of the physical components at $\nu_0$.
Therefore, the separation problem can be solved in terms of the hidden
vector ${\mathbf h}_{\ell m}$ and the corresponding ${\mathbf a}_{\ell m}$ are
subsequently found using (\ref{hidden}).
Note that ${\mathbf L}$ itself can be iteratively determined by the MEM (H98). 
That is, one can use an initial guess for ${\mathbf L}$ to obtain a first
reconstruction and subsequently compute the power spectra of those
reconstructions, which are used as a starting point for the next 
iteration, until convergence is achieved.

As shown in S02, harmonic-based MEM finds the best
reconstruction for the sky by minimising mode-by-mode the function
(for a detailed derivation see S02):
\begin{equation}
\Phi({\mathbf h}_{\ell m})=\chi^2({\mathbf h}_{\ell m})-\alpha
S({\mathbf h}_{\ell m},{\mathbf m}),
\label{phi}
\end{equation}
where $\chi^2$ is the standard misfit statistic in harmonic space
given by 
\begin{equation}
\chi^2({\mathbf h}\lm) = ({\mathbf d}\lm-{\mathbf R}_\ell {\mathbf L}_\ell
{\mathbf h}\lm)^\dag {\mathbf N}^{-1}_\ell 
({\mathbf d}\lm-{\mathbf R}_\ell {\mathbf L}_\ell {\mathbf h}\lm) 
\end{equation}
and $S({\mathbf h}_{\ell m},{\mathbf m})$ is the cross entropy (the
form of which is given in H98) 
of the complex vector ${\mathbf h}_{\ell m}$ and 
the model ${\mathbf m}$, to which ${\mathbf h}_{\ell m}$ defaults in 
absence of data. The regularising parameter $\alpha$ 
can be estimated in a Bayesian manner by
treating it as another parameter in the hypothesis space (see
$\S\ref{sec:alpha}$).

Working in harmonic space and neglecting correlations between
different modes vastly reduces the computational requirements of the
component separation problem . Instead of performing 
a single minimisation of $\sim 2n_c\ell_{max}^2$ parameters, one performs
$\sim \ell_{max}^2$ minimisations of $2n_c$ parameters, which is much
faster. In addition,
working in harmonic space provides us with a simple manner of
introducing (cross) power spectra information in the algorithm through
the $\mathbf{L}$ matrix.

\subsection{Flexible MEM}

Although the advantages of working in harmonic space are clear, there
are also some shortcomings. In addition to the necessity of neglecting
coupling between different $(\ell,m)$ modes, a full and regular
coverage of the sky is needed in order to keep the harmonic
transformations simple. The properties of the noise are also somewhat
diluted in harmonic space and it is not obvious how anisotropic or
correlated noise is affecting the algorithm. Finally, it is not
straightforward how to combine data taken in different spaces (such as
1D scans, interferometric data, incomplete spherical data, etc.).
Therefore, it would be desirable to develop a MEM algorithm that is
able to deal with all these issues, but still keeps as many advantages
of the harmonic-based MEM as possible. Unfortunately, this is a
non-trivial issue.

As a first approach, we have implemented a MEM algorithm that combines
the space where the data have been taken and harmonic
space. The best sky reconstruction is obtained by performing a
single minimisation, with respect to the whole set of parameters ${\mathbf h}
=\{{\mathbf h}\lm\}$, of the function 
\begin{eqnarray}
\mathbf{\varphi}({\mathbf h})&=&\chi^2_{\rm d}({\mathbf h})-\alpha
S({\mathbf h},{\mathbf m}),
\label{phireal}
\end{eqnarray}
where $\chi^2_{\rm d}$ is evaluated in the space where the data have been
taken. Note that, in this case, there is no need to transform the data
into harmonic space. Moreover, the properties of the noise are well
defined in data space and anisotropic noise can be easily included in
the analysis. This is simply achieved by including the relevant
noise dispersions in the noise covariance matrix of the misfit
statistics $\chi^2_d$. Also, incomplete sky coverage or galactic masks can be
taken into account, since the $\chi_{\rm d}^2$ can be calculated summing only
over a portion of the data. This method also allows the
combination of different set of data by combining their $\chi^2_{\rm d}$ 
values. The
cross entropy, however, is still calculated in harmonic space by summing over
the entropy $S({\mathbf h}_{\ell m},{\mathbf m})$ for each mode, thereby
preserving the straightforward introduction of power spectrum
information into the algorithm. The reconstruction is thus
still obtained in harmonic
space, making necessary an inverse transform from the harmonic modes to real
space (which is usually much simpler than the forward transform).
The form of $\chi_{\rm d}^2$, as a function of the hidden 
variables ${\mathbf h}\lm$, for spherical pixelised data is given in
Appendix~\ref{derivatives}.
Since we are not performing a mode-by-mode minimisation, correlations
between modes may be taken into account by the algorithm.

Unfortunately, the price that one has to pay for such a flexible
method is the need to perform a single minimisation of order $ \sim n_c
\ell_{max}^2$ parameters, which makes the method many times slower than
harmonic MEM. Even so, we believe that it is worth exploring the
possibilities of this algorithm, since it could be extremely
useful in many applications that can not currently be performed by
harmonic MEM.

\subsubsection{Newton-Raphson minimisation}

In order to perform the minimisation of the function given in
(\ref{phireal}), we use a Newton-Raphson (NR) iterative algorithm.
In the full NR method, the sky modes at iteration $i+1$ are obtained 
from their values at iteration $i$ via
\begin{equation}
{\mathbf h}^{i+1}={\mathbf h}^{i}-\gamma
({\mathbf H}^i)^{-1} {\mathbf g}^i,
\end{equation}
where `loop gain' $\gamma$ is a parameter 
of order unity, whose optimal value is
determined through the method itself, ${\mathbf g}^i$ is an 
$n-vector$ (where $n$
is the number of parameters to be 
minimised) containing the gradient of the function with respect to 
${\mathbf h}$ evaluated at ${\mathbf h}={\mathbf h}^i$, and ${\mathbf H}$ is
the curvature (or Hessian) matrix of dimension $n \times n$ 
evaluated at ${\mathbf h}^i$. The form of the first and second derivatives of 
$S$ are given in H98 and those of 
$\chi^2$ for spherical pixelised data are given in
Appendix~\ref{derivatives}. 

Unfortunately, with this scheme, one needs to calculate and invert
the Hessian matrix with dimension $n \times n$, which is not
feasible even for low-resolution data. Nevertheless, the main part
played by the Hessian matrix in the NR method is to provide a scale length
for the minimisation algorithm. Thus, the success of the algorithm does not
require a very accurate determination of the Hessian (indeed, many 
minimisation algorithms achieve very good results using only gradient 
information). In practice, therefore, we calculate only an
approximation to the Hessian matrix in the NR algorithm. We have found that
good results are obtained by approximating the
Hessian by a block diagonal matrix with blocks of size $n_c \times
n_c$, and setting the remaining elements to zero.
This corresponds to assuming that 
\[
\frac{\partial^2 h\lm^{k,c}}{\partial h_{\ell' m'}^{k',c'} 
\partial h_{\ell'' m''}^{k'',c''}} = 0 
\]
unless $(\ell,m)=(\ell',m')=(\ell'',m'')$ and $k=k'=k''$,
where the indices $k$ refers to the real or imaginary part of $h\lm$
and $c$ to each of the considered components to be reconstructed.
In this case, the inversion of the matrix can
be performed block-by-block, which simplifies the problem enormously.
Note that the value of each $h_{\ell m}$ mode still depends on the
values of all the rest of the modes through the gradient vector
(see Appendix~\ref{derivatives}). 

\subsubsection{Determining $\alpha$}
\label{sec:alpha}

Another important issue is the
determination of the regularisation parameter $\alpha$. As explained in H98, 
$\alpha$ can be obtained in a fully Bayesian manner by including it as
another parameter in the hypothesis space, and maximising the Bayesian
evidence with respect to it. In particular, one finds that it must satisfy
\begin{equation}
\alpha S(\hat{\mathbf h})=n - \alpha \mbox{Tr}({\mathbf M}^{-1}),
\end{equation}
where ${\mathbf M}={\mathbf G}^{-1/2}{\mathbf H}{\mathbf G}^{-1/2}$
and ${\mathbf G}$ is the (diagonal) metric of the image space (see H98
for details).  Note that in order to solve this implicit equation we
also need to operate with the Hessian matrix. To make this task
feasible, we again approximate $H$ by a block diagonal matrix with
blocks $n_c \times n_c$. This allows one to find a (nearly) optimal
value for the $\alpha$ parameter.

\subsubsection{Error estimation}

Following H98 and S02, we can also obtain an estimation of the 
covariance matrix of the
reconstruction errors $\widehat{\delta \mathbf a}_{\ell m}$ on the
harmonic modes $\hat{{\mathbf a}}\lm$, as well as 
the dispersion of the residual map for each component. 
In particular, by again approximating $H$ by a block
diagonal matrix, we find
\begin{equation}
\langle \widehat{\delta \mathbf a}\lm
\widehat{\delta \mathbf a}\lm^\dag \rangle=
\langle\left(\hat{\mathbf a}\lm- {\mathbf a}\lm\right)
\left(\hat{\mathbf a}\lm- {\mathbf a}\lm\right)^\dag \rangle=
{\mathbf L}_\ell {\mathbf H}\lm^{-1} {\mathbf L}_\ell^\dag
\end{equation}
where ${\mathbf H}\lm$ is the $n_c \times n_c$ block of the Hessian matrix
corresponding to the $(\ell,m)$ mode evaluated at $\hat{\mathbf
h}\lm$. As shown by S02, the former equation allows us to obtain
the residuals power spectrum at component $p$ simply by
\begin{equation}
\widehat{\delta C}_\ell^{(p)}
=\frac{1}{2\ell+1}\sum_{m=-\ell}^\ell 
\langle |\widehat{\delta {\mathbf a}}\lm^{(p)}|^2 \rangle
\end{equation}
where $\langle |\widehat{\delta {\mathbf a}}\lm^{(p)}|^2 \rangle$ is
given by the 
$p$th diagonal entry of the diagonal matrix ${\mathbf L}_\ell {\mathbf
H}\lm^{-1} {\mathbf L}_\ell^\dag$. Finally, we can estimate the
dispersion $e_{\rm est}$ of each residual map  as:
\begin{equation}
e_{\rm est}^2=\sum_{\ell} \frac{2\ell+1}{4 \pi} \widehat{\delta C}_\ell^{(p)}
\label{eq:e_est}
\end{equation}

\subsection{Estimating spectral behaviour}

In order to apply the MEM algorithm, one needs to assume that the
spectral behaviour of all the components to be reconstructed is known
and spatially constant. This is the case for the CMB and the SZ
effects, but it is not true for the diffuse galactic
components. Although there has been considerable effort in recent
years to study the galactic foregrounds at microwave frequencies,
there are still uncertainties in the knowledge of the frequency
dependence of free-free, synchrotron and dust emissions (as well as in
their spatial distribution). In addition, this frequency dependence is
expected to vary across the sky.  To study the synchrotron emission,
the most extensively used data has been the 408 MHz map of Haslam et
al. (1982). More recently, Reich \& Reich (1986) and Jonas et
al. (1998) measured the emission of the northern (at 1420 MHz) and
southern (at 2326 MHz) galactic hemispheres respectively. At these
frequencies the synchrotron emission dominates the total galactic
emission, and therefore these data sets are very useful in
characterising this foreground. The combination of these maps allows
one to estimate the frequency dependence of the synchrotron in this
frequency range. Reich \& Reich (1988) obtained an average value of
the synchrotron spectral index of -2.7 (assuming a power law in
temperature units) in the northern hemisphere. Giardino et al. (2002)
generated an spectral index map for the whole sky, whose average value
is also -2.7. More recently, the WMAP team (Bennett et al. 2003b)
found that the synchrotron power law in the WMAP range frequency is
relatively flat inside the galactic plane with an index of -2.5
whereas it steepens to $-3$ outside the Galaxy.

Thermal dust emission is usually modelled by a grey-body whose
emissivity depends on the physical properties of the materials that
constitute the dust (see e.g. Desert, Boulanger \& Puget, 1990; Banday
\& Wolfendale, 1991). Recently, there has been an effort to
produce a dust template, as well as to determine its frequency
dependence at microwave frequencies. In particular, Schlegel,
Finkbeiner \& Davis (1998) produced a dust map at 3000 GHz using IRAS
and COBE-DIRBE data. They also characterised the dust emission in the
1250-3000 GHz range with a grey body law with an emissivity of
$\alpha_d=2$ and a varying spatial temperature with values around 17-21
K.  More recently, Finkbeiner, Davis \& Schlegel (1999) proposed an
improved two-component dust model using the IRAS, COBE-DIRBE and
COBE-FIRAS data. The so-called cold component is characterised by a
mean temperature of 9.4 K and an emissivity of 1.67 whereas the
spectral parameters of the hot component take values of 16.2 K and
2.7 respectively.

The least well known of the galactic foregrounds is the free-free
emission. Usually, it is estimated through two different tracers: the
thermal dust emission and the H$_\alpha$ emission. Up to a few years
ago, no H$_\alpha$ surveys were available and thermal dust was
normally used to produce free-free templates, taking into account the
correlation found for instance between the COBE-DIRBE and COBE-DMR
data (Kogut et al. 1996). Recently, several H$_\alpha$ surveys have
been produced: the Virginia Tech Spectral line Survey (VTSS) of the
northern hemisphere (Dennison, Simonetti \& Topasna 1998), the
Wisconsin H-Alpha Mapper (WHAM) that covers a large fraction of the
sky (Reynolds, Haffner \& Madsen 2002) and the Southern H-Alpha Sky
Survey Atlas (SHASSA) of the southern hemisphere (Gaustad et
al. 2001). Even more recently, all-sky free-free templates have been
generated combining the information of the different surveys
(Dickinson, Davies \& Davis 2003, Finkbeiner 2003).  The
spectral dependence of the free-free emission is normally modelled as
a power law (in temperature) with spectral index around -2.16
(e.g. Kogut et al. 1996, Smoot 1998).

There are also uncertainties in the number of components
that contribute to the microwave sky. In fact, an anomalous galactic
emission at low frequency, which is well correlated with the thermal
dust one, has been found by several authors (de Oliveira-Costa et
al. 1997, Leitch et al. 1997 and Kogut 1999).  A possible candidate for this
component has been proposed by Draine \& Lazarian
(1998): electric dipole emission coming from rapidly rotating dust
grains (``spinning dust'').  The first statistical evidence for
spinning dust was given by de Oliveira-Costa et al. (1999). Finkbeiner
et al. (2002) found two tentative detections of this emission in two
(out of ten) small areas of the sky. However, Bennett et al. (2003b)
established that the spinning dust contribution to the WMAP
frequencies is clearly subdominant.

In summary, in order to apply our MEM algorithm, and taking into
account all the uncertainties in the knowledge of the galactic
foregrounds, we need a way to model their spectral behaviour, which
makes use of our prior knowledge of the foregrounds as well as of the data
themselves.  Additionally, it would be desirable to be able to
accommodate spatial variations of the spectral parameters.

One possibility to determine the (spatially constant) frequency
parameters from the data themselves would be to use an iterative
approach, assuming a known spectral law for the components. The
procedure would be as follows: (i) reconstruct the microwave
components with MEM using an initial guess for the unknown spectral
parameters; (ii) use the reconstructions as templates to fit for the
spectral parameters and the normalisation of the components (by
minimising $\chi^2$); (iii) use these new parameters as a starting
point to run MEM again; (iv) repeat the procedure until
convergence. We have tested this method on our set of simulated data,
but unfortunately it did not always converge to the correct
values. However, these data have a low signal-to-noise ratio and we
need to fit for several spectral parameters at the same time, which
leads to degeneracies. In addition, MEM will drive the reconstructions
towards the templates that best fit the initial (incorrect) spectral
parameters, which may be far from the templates that fit the correct
parameters. All these factors make it very difficult to estimate the
frequency dependence of the components with this method for our set of
data. Nevertheless, if high quality data are available, or if a single
component needs to be fitted, this method should be further
investigated.

\begin{figure*}
  \begin{center}
    \includegraphics[width=16cm]{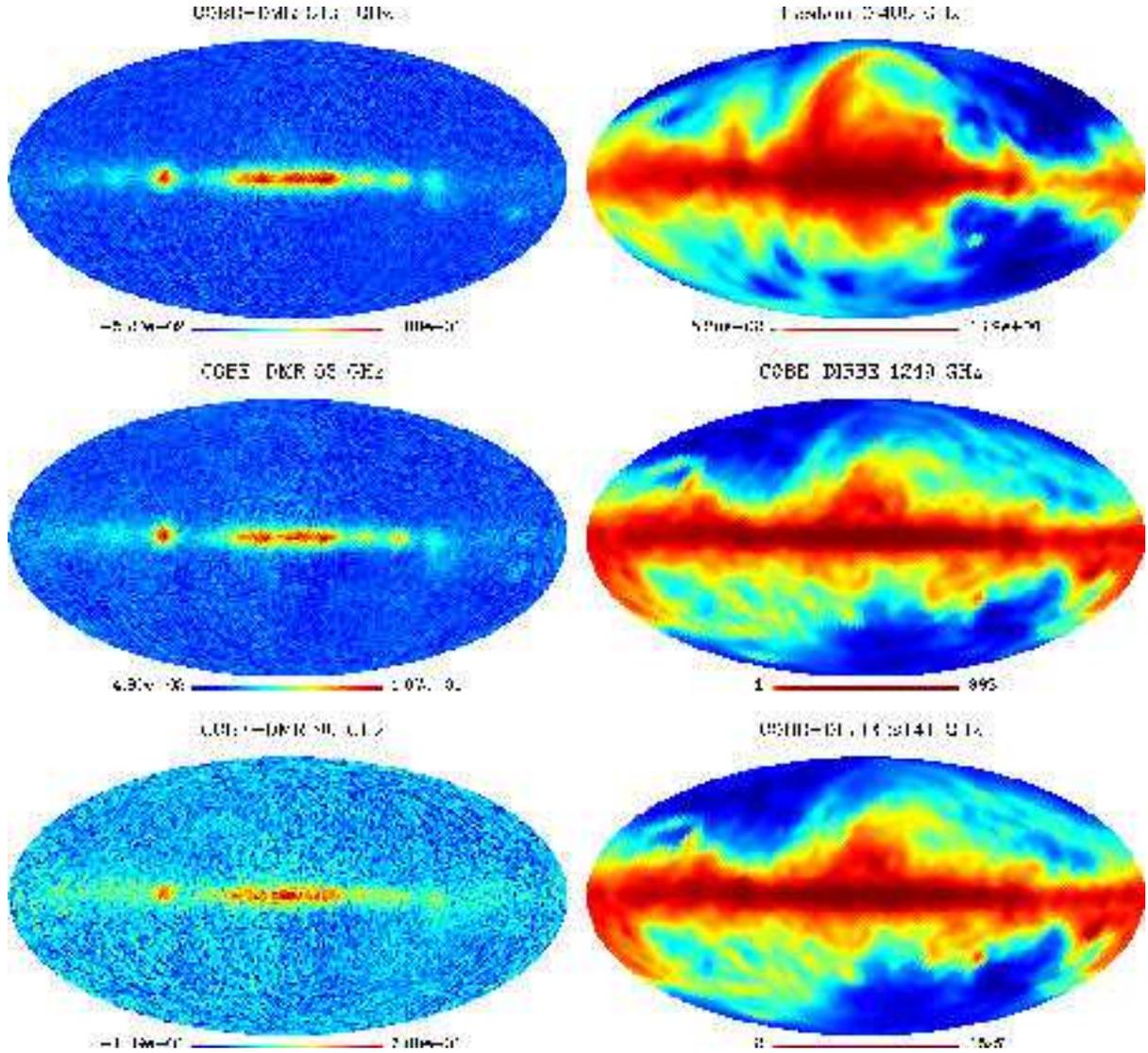}
  \end{center}
  \caption{\label{fig:data} Data maps used in our analysis in units of
    MJy/sr. The left column shows the three COBE-DMR channels
    (from top to bottom): 31.5, 53 and 90 GHz. The top right panel
    gives the Haslam map and the middle and bottom right panels
    correspond to
    two of the COBE-DIRBE channels (1249 and 2141 GHz). The Haslam and
    COBE-DIRBE maps have been plotted in a non-linear scale to allow
    the visualization of the structure of the maps outside the
    galactic centre}
\end{figure*}

A different approach is to run MEM for
different sets of spectral parameters and try to infer from the
reconstructed components which parameters give the best results and
are, therefore, closest to the truth. 
One could naively look at the estimated errors and pick the
reconstructions with the lowest values for these errors. However, the
estimation of the errors depend on the chosen spectral
parameters. Basically, it gives the statistical error of the
reconstruction but it does not take into account uncertainties in the
values of the spectral parameters.
Therefore, if our guess of the frequency dependence is
incorrect, the error estimation of the reconstructions is not reliable.
Jones et al. (2000) showed that the dust frequency dependence could be
estimated from Planck simulated data of small patches of the sky by
minimising the $\chi^2$ of the reconstructions. They assumed a
(spatially constant) grey body law for the
dust emission with two unknown parameters (the dust temperature and
emissivity), which they were able to fit from the data by looking at the
minimum of the $\chi^2$ of the reconstructions. However, varying the
spectral index $\beta$ of the synchrotron or free-free emissions (assuming
a power law $I\propto \nu^{\beta}$) had little effect on the value 
of the $\chi^2$ and could not be determined in this way. This was due
to the fact that both the synchrotron and free-free emissions have a
low amplitude with respect to the rest of the components in the
considered region of the sky and therefore the data could not provide
enough information to fit these two components.
In this case, the reconstructions of these two emissions were lost, but
those of the other components were little affected.
This application shows that, depending on the characteristics of the
data (sky coverage, resolution, signal-to-noise ratio, frequency
coverage, etc.) we may not be able to estimate some of the spectral
parameters using just the $\chi^2$ value.
Therefore, in order to estimate all the spectral parameters we may need 
to use also information coming from other variables, such as the
entropy or the cross correlations between the reconstructed CMB and
the galactic components. We have used such an approach to determine
empirically the best reconstructions, which is explained
in detail in $\S\ref{rec:goodness}$. 

\section{Spherical microwave datasets}
\label{sec:data}

The only spherical CMB dataset available during the development of
the flexible MEM algorithm was COBE-DMR data. In order to enable the
reconstruction of different components of emission, we also make use
of the COBE-DIRBE and Haslam maps. Each of these maps exist in
the HEALPix pixelization (G\'orski et al. 1999) with $N_{\rm side}=32$, which
corresponds to 12288 pixels of size 107 arcminutes.
Since each of these datasets has been described in detail elsewhere,
we just give a brief description in this section.
The data maps in MJy/sr are shown in Fig.~\ref{fig:data}.

\begin{figure}
  \begin{center}
    \includegraphics[width=8cm]{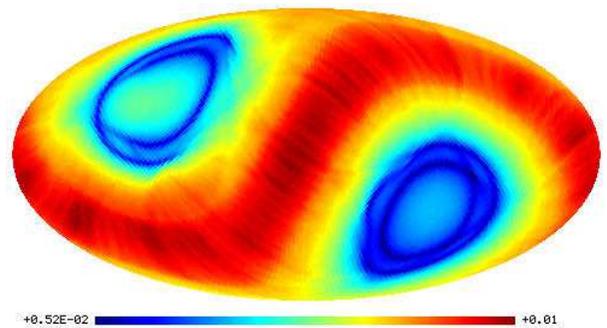}
  \end{center}
  \caption{\label{fig:noise53} Dispersion noise per pixel of the
  COBE-DMR 53 GHz map in MJy/sr.}
\end{figure}

\begin{figure}
\includegraphics[angle=-90, width=8cm] {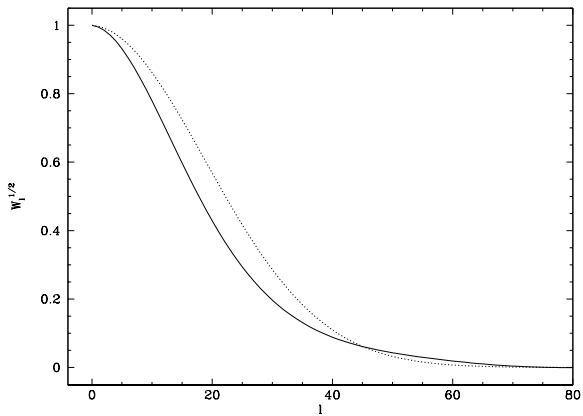}
\caption{Window function of the COBE-DMR instrument (solid line). For
comparison a Gaussian beam of FWHM=7$^{\circ}$ is also shown (dotted line).}
\label{fig:beam}
\end{figure}

The COBE-DMR data consists of three frequency maps: 31.5, 53 and 90
GHz (each of them obtained by the combination of two channels) with a
resolution of $\sim 7$ degrees and a signal-to-noise ratio of 
around 2 per 10 degree patch. 
The COBE-DMR data have an anisotropic noise pattern that can 
be easily taken into account with our method. As an illustration we show the
noise dispersion per pixel for the 53 GHz channel in
Fig~\ref{fig:noise53}. The COBE beam is well characterized by the
curve given in Fig.~\ref{fig:beam} (Wright et al. 1994). 
For comparison, a Gaussian beam with 7
degree of full width half maximum (FWHM) is also shown.

The COBE-DIRBE experiment observed the brightness of the full sky
at ten wavelengths from 1.2 to 240 microns as well as mapping linear
polarization at 1.2, 2.2 and 3.5 microns. In our analyses below we use 
the data at the two lowest frequencies (1249 and 2141 GHz).
The maps have been degraded down to $N_{side}=32$ and smoothed
with a Gaussian beam of FWHM 2.4 times the pixel size (i.e., FWHM
=263.8 arcmin). The COBE-DIRBE noise is
also anisotropic, but since we have significantly degraded the
resolution of the data, its level is very small.
The COBE-DIRBE beam has a resolution of 0.7 degrees and can
be approximately modelled by a top-hat.
At these frequencies, (thermal) dust emission dominates and therefore
these data help the algorithm to extract the dust component.

The Haslam map gives the emission of the whole sky at 408 MHz and has
been obtained  by combining four different surveys (Haslam et
al. 1982). It has an effective resolution of 51 arcminutes. 
Recently, Finkbeiner, Davis \& Schlegel (private communication) 
have reprocessed the original Haslam map to provide a destriped,
point-source subtracted map, which we have used for our analysis.
We have degraded its resolution down to the HEALPix resolution of 
$N_{side}=32$ and smoothed it with a Gaussian beam of FWHM=263.8 arcmin. 
At this frequency the emission is dominated by
galactic synchrotron, providing MEM with excellent
information to trace this component.

\section{Analysis of simulated data}
\label{sec:results_simu}

\begin{figure*}
  \begin{center}



    \includegraphics[width=16cm]{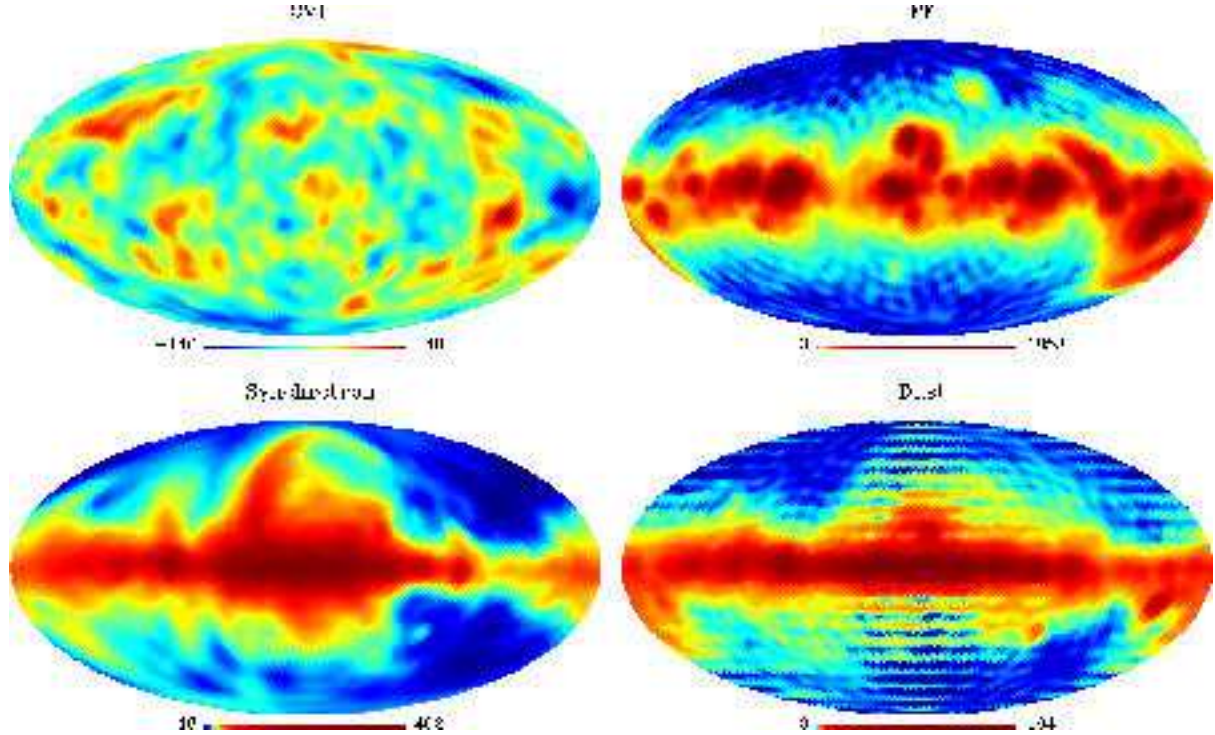}

  \end{center}
  \caption{\label{fig:components} Input simulated components (from
  left to right and top
  to bottom): CMB, free-free, Synchrotron and dust at 50
  GHz in $\mu$K (thermodynamical temperature). The maps have been
  smoothed with a 7 degree Gaussian 
  beam. The galactic components have been plotted in a non-linear
  scale to allow for a better visualization of the high-galactic
  latitude region. To allow for a straightforward comparison with the
  reconstructed map, the input components include only information up
  to $\ell=40$. This produces the ringing artifacts visible in the dust map.}
\end{figure*}

Before applying of our flexible MEM component separation method to 
real data, we have checked its performance on simulated datasets.
In order to simulate our sets of data (COBE-DMR, COBE-DIRBE and Haslam
maps) we have assumed that the sky contains, CMB, free-free,
synchrotron  and thermal dust emissions.
The four simulated maps at the reference frequency of 50 GHz are given
in Fig.~\ref{fig:components} in units of $\mu$K (thermodynamical
temperature). 

\subsection{The simulations}

The CMB realization has been produced with the help of the HEALPix
package and has been constrained to have a power
spectra compatible to the one derived from the COBE-DMR data 
(Bennett et al. 1996) at low $\ell$'s and
from Archeops (Benoit et al. 2003) for higher multipoles.
The synchrotron emission has been modelled by a power law $I_{\rm syn}
\propto \nu^{-0.8}$ and the template at the reference frequency has
been obtained by extrapolating the Haslam map with this law.
The dust emission has been obtained by extrapolating the IRAS-DIRBE
map\footnote[1]{This map is available at http://skymaps.info} 
of Schlegel, Finkbeiner \& Davis (1998) using a single grey
body component with dust emissivity of 2 and temperature of 18 K.
Finally, the free-free component has been modelled with a power law of 
$I_{\rm ff} \propto \nu^{-0.16}$. As template for the free-free we
have used the 
WHAM survey (Reynolds, Haffner \& Madsen 2002), which provides an 
H$_\alpha$ map of most of
the sky. The empty part of the sky has been filled in by copying on it
another region of the survey. The map has been normalized to have a
dispersion of 185 $\mu$K at 50 GHz and smoothed with a 7 degree
Gaussian beam. This is several times higher than expected using
estimations from the H$_\alpha$ emission (e.g. Finkbeiner 2003), 
but it is intended to account for the
excess of emission found in the data whose origin is uncertain.

\begin{table}
\begin{center}
\caption{Assumed observational parameters for the simulated data}
\label{tab:data}
\begin{tabular}{ccccccc}
\hline
& Haslam & \multicolumn{3}{c}{DMR} & \multicolumn{2}{c}{DIRBE}\\
\hline
Freq (GHz) & 0.408 & 31.5 & 53 & 90 & 1249 & 2141 \\
Resolution ($^\circ$) & 4.5 & $\sim$ 7 & $\sim$ 7 & $\sim$ 7 & 4.4 & 4.4 \\
$\Delta \nu / \nu$ & 0.04 & 0.07 & 0.04 & 0.05 & 0.28 & 0.40 \\
\hline
\end{tabular}
\end{center}
\end{table}

The separate components have been combined in order to simulate data
according to the characteristics given in
Table~\ref{tab:data}. For the COBE-DMR data we have used the true
beam shape given in Fig.~$\ref{fig:beam}$ and
added Gaussian pixel noise 
according with the anisotropic pattern of the maps. 
Since we aim to reconstruct only four components (and in particular a
single dust component) we have simulated only the lowest frequency 
COBE-DIRBE channel. We have ignored the beam of this experiment 
since it is small in  
relation to the pixel (and the subsequent smoothing). 
We have added the corresponding anisotropic noise at the COBE-DIRBE 
map. Signal plus noise
have then been smoothed with a Gaussian beam of 263.8 arcmin. 
Finally, since the real Haslam map has been
produced combining different surveys and has been degraded and
reprocessed, it is not straightforward to determine the level of pixel
noise present in the map, which will also be correlated. 
However, we expect it to be very small for the
considered scales. Therefore, we have neglected the noise when generating the
map, which has been simulated with an effective resolution of 268.7
arcminutes (51 arcminutes beam coming from the resolution of the Haslam map
plus a smoothing of FWHM=263.8 arcminutes). 

\subsection{Results}

We have applied the method explained in $\S\ref{sec:method}$ to our
simulated data in order to recover the CMB and galactic foregrounds.
Given the resolution and signal to noise of the data, we have aimed to
reconstruct the different components only up to $\ell_{max}=40$, since
there is virtually no information about the CMB in the data at higher
multipoles. We need to provide the algorithm with an initial guess
for the power spectra for each of the components. As shown in H98, 
the reconstructions 
are not very sensitive to this initial guess, provided one iterates 
over the power
spectra, i.e. one performs the reconstruction with an initial guess and
uses those reconstructed maps to provide starting power spectra for
the next iteration, until convergence is obtained. 
For the galactic components, we have chosen initial power spectra
that differ appreciably from the original input maps to show the
performance of the method even when the initial power spectra are far
from the correct ones. Regarding the CMB, we have chosen not to
iterate over its 
power spectrum, but to start each iteration with a 
CMB model which is compatible (but that differs from the input one) 
with the power spectra derived from the COBE data (at the
lowest $\ell$'s) and from Archeops (at the highest multipoles).
As shown in S02, an approximated initial guess is enough for MEM
to find the underlying power spectra. 
It must also be pointed out that if we were using a prior for the CMB
power spectrum that significantly differs from the true one and we do 
not iterate over this quantity, this could bias the results
obtained by the method, especially at those scales with a low signal
to noise ratio. However, we do have a good knowledge of the
shape of the power spectrum of the CMB at these low
$\ell$'s. Therefore it is reasonable to make use of these information
in order to improve the CMB reconstruction.
In any case, we would like to emphasize that the knowledge of the CMB (or any
of the other components) power spectrum is not necessary for the
method to work. If there was a total absence of knowledge of the CMB
prior, we would need to iterate over its power spectrum to find the
correct reconstruction. In fact, if we were using high quality data,
such an analysis without providing any prior information should also
be performed. This would avoid biasing the results as well as
point out possible inconsistencies with the supplied prior.

We also need an estimation of the noise for each data map. The noise
of the COBE maps is well known, but this is not the case for the
Haslam map, as mentioned in the previous section. In order to provide
a reasonable value for the calculation of the $\chi^2$ function we
have used the following trick. 
The Haslam map has some structure
beyond $\ell=40$, but we are going to reconstruct the map only up
to that multipole. Since the map has been smoothed and processed we
expect the noise to be very small at the scales that we are
reconstructing. Therefore when
subtracting our predicted data map
(generated using the reconstructions up to $\ell_{max}$) from
the true data map to obtain the $\chi^2$, the difference will come
mainly from the power beyond the maximum reconstructed multipole and
that can be considered our \emph{effective} noise. Therefore, the estimation
of the noise for this simulated data map has been obtained as the
dispersion of 
the map obtained subtracting the simulated Haslam map with power up to
$\ell_{max}$ from the same map at full resolution.
In practice we have used the same trick to estimate the noise of the
COBE-DIRBE channel. This is due to the fact that the noise per pixel
of this map is very low after repixelisation and smoothing and it 
was necessary to take into account the structure present beyond
$\ell_{max}$, which was giving the main contribution to the $\chi^2$.
This estimation of the noise produced $\chi^2$ in the correct range.

\subsubsection{Estimation of the spectral parameters in a low noise case}
\label{sec:lownoise}
Unfortunately the COBE-DMR data maps are very noisy, so it is
difficult to test the performance of our method and, in particular, the
determination of the spectral parameters, with this data
set. Therefore, we have first tested our flexible MEM algorithm in an
artificially low noise case. For this test, we have used the same
simulations described in the previous section but the noise level of
the COBE-DMR channels has been lowered by a factor of 5.

We have applied our method to this simulated data assuming different sets of
spectral parameters. In particular, we have used all the possible
combinations (a total of 81) of the following values: $\beta_{\rm
ff}$=-0.13,-0.16,-0.19, $\beta_{syn}=-0.7,-0.8,-0.9$, $T_{\rm
d}$=16,18,20 and $\alpha_{\rm d}$=1.8,2.0,2.2. 
For each set of spectral parameters we have iterated over the
power spectra for all the galactic components to find the best
possible reconstructions in each case.
We have then looked at the $\chi^2$ value of the reconstructed maps
for each combination of spectral 
parameters as an indicator of the quality of the
reconstructions. We find that, as one would expect in an ideal case,
the reconstruction with the lowest $\chi^2$ value (or equivalently
$\chi^2/n_f$) corresponds to the one
obtained using the correct set of spectral paramaters (i.e., $\beta_{\rm
ff}$=-0.16, $\beta_{syn}=-0.8$, $T_{\rm d}$=18 and $\alpha_{\rm
d}$=2.0). Since the noise of each data map is quite small, there is no
room for uncertainties in the spectral indices and the $\chi^2$
successfully picks the right set of spectral parameters.
We have numbered the different cases according to the obtained
$\chi^2$ value. Case number 1 corresponds to the lowest $\chi^2$ whereas case
81 is the one with the highest $\chi^2$
value. Table~\ref{tab:best_simu_low} gives  
the value of the spectral parameters and of $\chi^2/n_f$ for the 10 first
cases. 
\begin{table}
\begin{center}
\caption{Spectral parameters and value of $\chi^2/n_f$ for the
ten best cases 
from low noise simulated data}
\label{tab:best_simu_low}
\begin{tabular}{c|ccccc}
\hline
Case & $T_d$ & $\alpha_d$ & $\beta_{ff}$ & $\beta_{syn}$ & $\chi^2/n_f$ \\
\hline
1 & 18 & 2.0 & $-0.16$ & $-0.8$ & 12156 \\
2 & 18 & 2.0 & $-0.19$ & $-0.8$ & 12167 \\
3 & 18 & 2.0 & $-0.13$ & $-0.8$ & 12171 \\
4 & 18 & 2.0 & $-0.19$ & $-0.9$ & 12215 \\
5 & 16 & 2.2 & $-0.13$ & $-0.8$ & 12229 \\
6 & 16 & 2.2 & $-0.16$ & $-0.8$ & 12237 \\
7 & 16 & 2.2 & $-0.19$ & $-0.9$ & 12240 \\
8 & 18 & 2.0 & $-0.16$ & $-0.9$ & 12244 \\ 
9 & 20 & 2.0 & $-0.13$ & $-0.8$ & 12252 \\
10& 20 & 2.0 & $-0.16$ & $-0.8$ & 12258 \\
\hline
\end{tabular}
\end{center}
\end{table}

\begin{figure*}
\includegraphics[width=16cm] {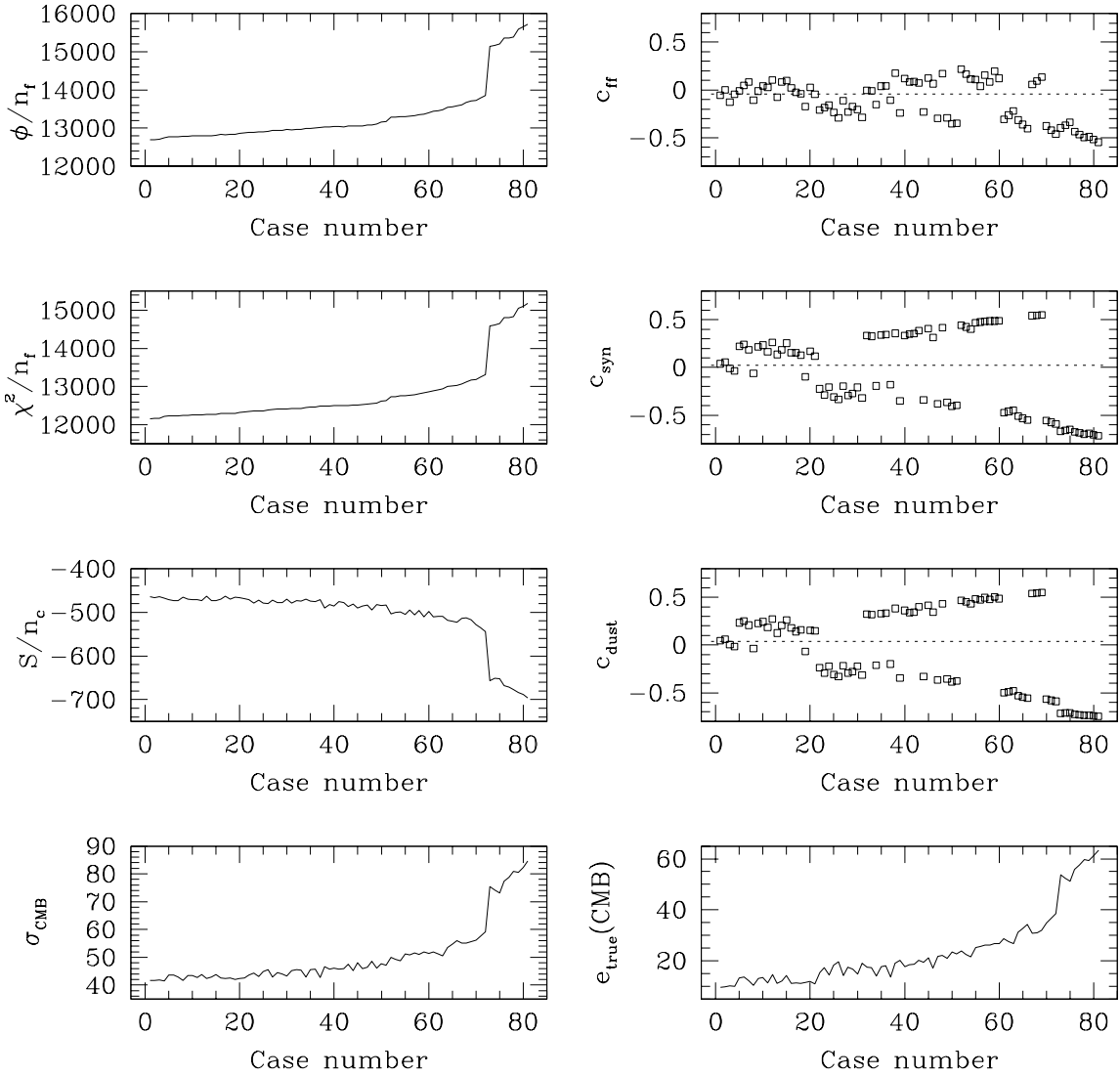}
\caption{The following quantities have been plotted versus the case
number for the low-noise simulated case:
left column (from top to
bottom): $\varphi/n_f$, $\chi^2/n_f$, $S/n_c$ and dispersion of the
reconstructed CMB map. Right column (from top to bottom): cross
correlations between the reconstructed maps of CMB and free-free, CMB and
synchrotron, CMB and dust and error of the CMB reconstruction. For reference, 
the dotted line in the cross correlation panels show the level of the
correlation between the input maps of the CMB and the corresponding galactic
component.} 
\label{fig:chi2_simu_low}
\end{figure*}

In addition to the $\chi^2$ there are other quantities that can be
calculated from the reconstructed components that also provide
information about the quality of the reconstructions. In particular,
we have also studied the behaviour of $\varphi/n_f$, $S/n_c$, the
cross-correlations between the CMB reconstructed map and the three galactic
components (which should be zero) and the dispersion of the CMB
reconstructed map $\sigma^{\rm rec}_{\rm CMB}$. All these quantities are plotted in
Fig.~\ref{fig:chi2_simu_low} versus the case number. The error
$e_{\rm true}$ in the CMB reconstruction (smoothed with a 7 degree beam)
is also shown and has been calculated as
\begin{equation}
e_{\rm true}=\sqrt{\langle (T_i - T_r)^2 \rangle - \langle T_i - T_r\rangle ^2}
\end{equation}
where $T_i$ and $T_r$ correspond to the input and reconstructed map
smoothed with a 7 degree beam.
It is interesting to note the high correlation
between the value of the $\chi^2$ and the CMB reconstruction
error. Moreover, this error correlates also very strongly with the
rest of the plotted quantities. As one would expect, the maps with a
low $e_{\rm true}$ also present values of the cross-correlations between
the CMB and the galactic restored maps close to zero. Small values of
$\varphi$ 
and $|S|$ correspond as well to small values of $e_{\rm true}$. Finally, we
also find that the lowest values of $\sigma^{\rm rec}_{\rm CMB}$ are
also the ones with 
better CMB reconstructions. This can be explained taking into account
that those CMB recostructions obtained with wrong spectral indices
will be contaminated by galactic emission and therefore the dispersion
of the CMB map will increase.

Since the main objective of this test was to study the estimation of 
the spectral parameters using different indicators of the
quality of the reconstructions, we will not go into detail with regard
to the reconstructed maps and power spectra. However we give, as
reference, the difference reconstructed errors for the four recovered
components in Table~\ref{tab:rms_simu_low} for case 1.
\begin{table}
\begin{center}
\caption{Dispersion values for input, reconstructed (from case 1) 
and residuals maps smoothed with
a 7 degree beam at 50 GHz in $\mu$K for the low noise simulated
data. These values 
are given for the whole map (col.~3) , for the region outside the
galactic cut (col.~4) and for the galactic centre (col.~5). 
For comparison the estimated error is also given in the last column.}
\label{tab:rms_simu_low}
\begin{tabular}{c|ccccc}
\hline
Cpt & Map & $\sigma_{all}$ & $\sigma_{out}$ & $\sigma_{Gal}$  & $e_{est}$\\
\hline
& Input & 36.4 & 36.5 & 36.3 & \\
CMB & Rec. & 35.8 & 35.4 & 36.5 & \\
& Resid. & 9.5 & 9.4 & 9.8 & 10.9\\
\hline
& Input & 186.2 & 34.8 & 255.6 & \\
FF & Rec. & 186.6 & 36.5 & 256.2 & \\
& Resid. & 11.9 & 11.8 & 11.9 & 23.7\\
\hline
& Input & 42.7 & 10.1 & 59.2 & \\
Synch. & Rec. & 42.7 & 10.1 & 59.2 & \\
& Resid. & 0.55 & 0.54 & 0.55 & 1.4\\
\hline
& Input & 16.6 & 0.66 & 25.1 & \\
Dust & Rec. & 16.6 & 0.66 & 25.1 & \\
& Resid. & 0.08 & 0.08 & 0.08 & 1.3\\
\hline
\end{tabular}
\end{center}
\end{table}

\subsubsection{Estimation of the spectral parameters in the 
simulated case with realistic noise}
\label{rec:goodness}
\begin{figure*}
\includegraphics[angle=-90, width=16cm] {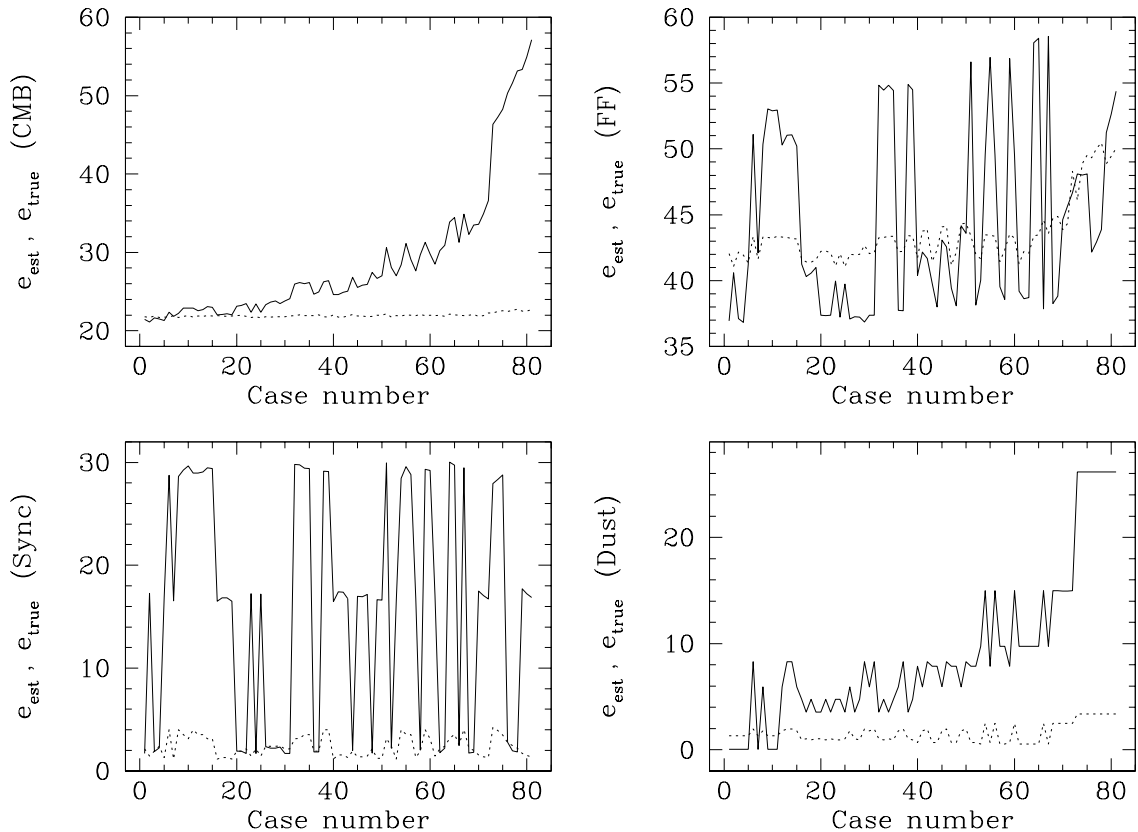}
\caption{The panels show, for the realistic simulated data,
the residuals dispersion (solid line) versus the
case number for each of
the reconstructed components after smoothing with a 7 degree Gaussian
beam. For comparison the estimated error is also plotted (dashed line)}
\label{fig:errors_simu}
\end{figure*}
\begin{figure*}
\includegraphics[width=16cm] {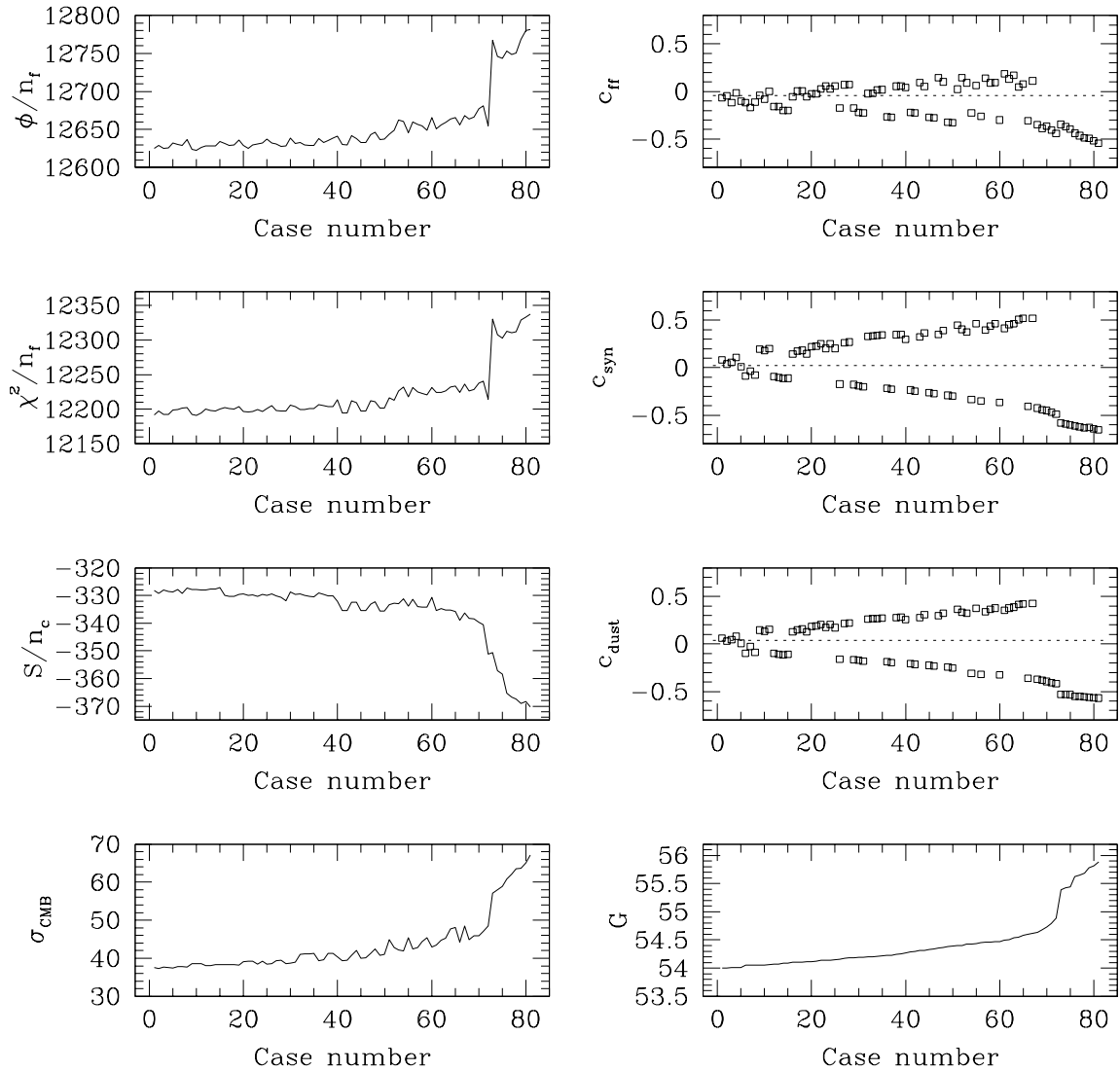}
\caption{The quantities used to calculate $G$, and $G$
itself, are plotted versus the case number for the simulations. 
Left column (from top to
bottom): $\varphi/n_f$, $\chi^2/n_f$, $S/n_c$ and dispersion of the
reconstructed CMB map. Right column (from top to bottom): cross
correlations between the reconstructed maps of CMB and free-free, CMB and
synchrotron, CMB and dust and $G$. For reference, 
the dotted line in the cross correlation panels show the level of the
correlation between the input maps of the CMB and the corresponding galactic
component.} 
\label{fig:chi2_simu}
\end{figure*}
\begin{figure}
\includegraphics[width=8cm] {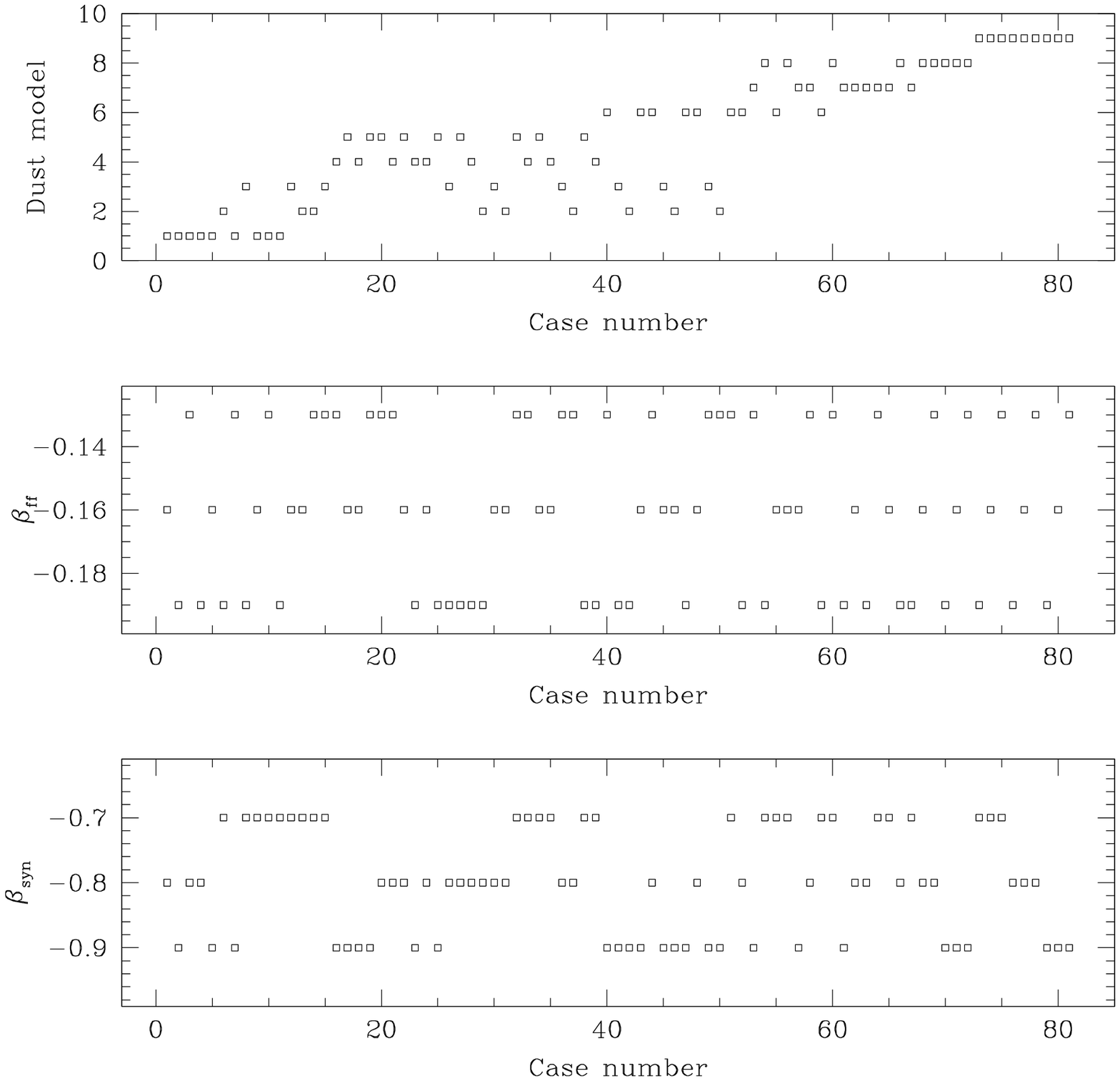}
\caption{Spectral parameters versus case number for the
simulations: dust model (top), spectral index for the free-free (middle) and
spectral index for synchrotron (bottom). The dust model number correspond
to: $T_d$=18,$\alpha_d$=2 (1), $T_d$=20,$\alpha_d$=1.8 (2),
$T_d$=16,$\alpha_d$=2 (3), $T_d$=16,$\alpha_d$=2.2 (4), 
$T_d$=20,$\alpha_d$=2 (5), $T_d$=18,$\alpha_d$=2.2 (6), 
$T_d$=20,$\alpha_d$=2.2 (7), $T_d$=18,$\alpha_d$=1.8 (8), 
$T_d$=16,$\alpha_d$=1.8 (9).}
\label{fig:param_simu}
\end{figure}
We have repeated the previous analysis using our (realistic)
simulated data set and have obtained reconstructions for the same 81
combinations of spectral parameters.
As in the previous case the reconstructions have been obtained
iterating over the power spectra of the galactic components for each
set of spectral parameters.
We have then investigated the behaviour of the
different quantities that might indicate which combination of
parameters is correct. 
If we look at the reconstruction with the lowest $\chi^2$ value we
find the following set of parameters: $T_d=18$, $\alpha_d=2.0$,
$\beta_{\rm ff}=-0.13$, $\beta_{\rm syn}=-0.7$ (corresponding to a
$\chi^2/n_f=12191$) whereas the reconstruction with the correct
spectral parameters has the second lowest $\chi^2$ value which is also
very close to the minimum ($\chi^2/n_f=12192$). In fact, 3 more of the
reconstructions have values $ \chi^2/n_f < 12193$. Therefore we find
that the data are just too noisy to discriminate clearly between
different spectral data sets using just the
information given by the $\chi^2$. However, we have seen in the
previous section that there is more information that we can extract
from the reconstructed maps to assess the quality of the
reconstructions and to determine the spectral parameters.
Therefore, we have also considered the
values of the entropy $S$, the $\varphi$ function, the dispersion of the
reconstructed CMB map $\sigma_{\rm CMB}^{\rm rec}$, and the cross-correlation 
between the reconstructed CMB and each of the reconstructed galactic
components. 
In particular we have constructed an empirical selection function
$G$ that is a linear combination of the former quantities
and is defined as
\begin{eqnarray}
G & = & \frac{1}{\sum_{i} a_i}\left[a_1 \frac{\chi^2}{n_f} + a_2
\frac{\varphi}{n_f} + a_3 \frac{|S|}{n_c} 
+ a_4 |c_{\rm ff}| \right. \nonumber \\
& & \hspace{10mm}\left. \phantom{\frac{\chi^2}{n_f}} + a_5 
|c_{\rm syn}| + a_6 |c_{\rm dust}| + a_7 \sigma_{\rm CMB}^{\rm rec}\right],
\end{eqnarray}
where $c_j$ denotes the cross-correlation between the CMB
reconstruction and the corresponding galactic component.
The minimum of $G$ will give us the best reconstructions. The chosen
weights $a_i$ are given in
Table~\ref{tab:weights} and have been determined by looking for an
optimal combination that gives more weight to those reconstructions
with higher quality and whose spectral parameters deviate less from
the true ones. The goal of combining all this information is that
even if we can not reliably determine the spectral parameters, given
the low signal to noise of the data, at least we can 
prevent the uncertainties in the knowledge of such parameters from
severely contaminating the CMB reconstruction.

\begin{table}
\begin{center}
\caption{Weights for the calculation of $G$}
\label{tab:weights}
\begin{tabular}{c|c}
\hline
Coefficient   & Value\\
\hline
$a_1$ & 0.6 \\
$a_2$ & 0.6 \\
$a_3$ & 5 \\
$a_4$ & 100 \\
$a_5$ & 100 \\
$a_6$ & 100 \\
$a_7$ & 1 \\
\hline
\end{tabular}
\end{center}
\end{table}
\begin{table}
\begin{center}
\caption{Spectral parameters and value of $G$ for the ten best cases
from (realistic) simulated data}
\label{tab:best_simu}
\begin{tabular}{c|ccccc}
\hline
Case & $T_d$ & $\alpha_d$ & $\beta_{ff}$ & $\beta_{syn}$ & $G$ \\
\hline
1 & 18 & 2.0 & $-0.16$ & $-0.8$ & 54.0050 \\
2 & 18 & 2.0 & $-0.19$ & $-0.9$ & 54.0055 \\
3 & 18 & 2.0 & $-0.13$ & $-0.8$ & 54.0063 \\
4 & 18 & 2.0 & $-0.19$ & $-0.8$ & 54.0084 \\
5 & 18 & 2.0 & $-0.16$ & $-0.9$ & 54.0085 \\
6 & 20 & 1.8 & $-0.19$ & $-0.7$ & 54.0525 \\
7 & 18 & 2.0 & $-0.13$ & $-0.9$ & 54.0536 \\
8 & 16 & 2.0 & $-0.19$ & $-0.7$ & 54.0547 \\
9 & 18 & 2.0 & $-0.16$ & $-0.7$ & 54.0556 \\
10& 18 & 2.0 & $-0.13$ & $-0.7$ & 54.0563 \\
\hline
\end{tabular}
\end{center}
\end{table}

Using $G$, we find that the preferred reconstruction is the one with the
correct spectral paramaters, whereas that with the lowest $\chi^2$
value is now in position 10 (see Table~\ref{tab:best_simu} to see the
values of the parameters for the first 10 cases). The different cases
have been ordered according to the value of $G$, so those cases with
a lower number correspond to sets of parameters that produce a smaller $G$.
It is interesting to note that the smoothed 7 degree CMB
reconstruction error of the 
case prefered by the $G$ quantity (case 1) is $21.5\mu$K whereas that
of the case with the minimum $\chi^2$ (case 10) is $22.9\mu$K. In fact
if we calculate the correlation between the values of $G$ and of the CMB
reconstructed error for the 81 cases we find a value of 0.993 versus
0.969 for the correlation between the $\chi^2$ value and the same
error. In addition, there is also another importan reason to
use the combined information rather than just the $\chi^2$. 
For real data, where the foregrounds can not be well
modelled by a simple law, or where systematics may be present, it seems a
good idea to combine all the available information.
In particular, the $\chi^2$ will not perform as well as
in the ideal case and therefore we should also make use of quantities such
as the cross-correlations between CMB and the galactic components that
can be more reliable to assess the quality of the reconstruction for
real data. 

The behaviour of the quantities used to construct $G$ 
is illustrated in Figs.~\ref{fig:errors_simu},
~\ref{fig:chi2_simu} and ~\ref{fig:param_simu}. The different
quantities have been plotted versus the case number (ordered according
to $G$); Fig.~\ref{fig:errors_simu} shows the
dispersion of the residuals (solid line) for each of the components smoothed
with a 7 degrees Gaussian beam. Note that in the
CMB case, there is a clear correlation between the reconstruction
error and the 
ordering of the cases: a lower value of $G$ (which is plotted in 
the bottom right panel of Fig.~\ref{fig:chi2_simu}) will produce in general a
better CMB reconstruction. It is also striking that this error
varies very little for the cases with the lowest values of $G$
In fact, the difference outside the galactic cut\footnote[2]{We will
always refer to the 
custom galactic cut of Banday et al.(1997) in HEALPix 
pixelization, which masks a total of 4594 pixels.} 
between the CMB reconstruction (smoothed with a
7 degree Gaussian beam) of case 1 and
all those up to case 25 is $\le 3 \mu$K, 
which is well below the statistical errors. As expected, the CMB
reconstruction in the galactic region is more dependent on the
spectral parameters and the differences range between $3-12 \mu$K for
the same cases. The dashed lines in Fig.~\ref{fig:errors_simu} indicate the 
residuals dispersion estimated by our method. As already mentioned,
this error is 
only reliable if the spectral parameters are close to the true
values. If this is the case, this error gives a good estimation of the
residuals dispersion. 

Fig.~\ref{fig:chi2_simu} shows some additional sensitive quantities versus the
case ordering. We see 
that low values of $\chi^2$, $\varphi$ and of the absolute
value of the entropy also go in the direction of producing better
reconstructions. In addition, the cross correlations between the
reconstructed CMB and galactic components are also indicators of the
reliability of the reconstructions.
Another very interesting result is given in the bottom
left panel of Fig.~\ref{fig:chi2_simu}, where the dispersion of the
reconstructed CMB map smoothed with a 7 degree Gaussian beam is given.
The correlation between this quantity and the actual error of the CMB
map is striking. As already mentioned, this can be understood since
errors in the CMB 
restored map would mainly come from the introduction in the
reconstruction of galactic contamination, which will give rise to a
higher dispersion of the map.
The fact that different foreground models produce similar CMB
reconstructions indicate that we have degenerate cases, due again to
the low signal-to-noise ratio of our data.

This effect can be seen in the two bottom panels of
Fig.~\ref{fig:param_simu}, where the spectral indices for the free-free and
synchrotron are plotted versus the case number. There is no  
clear trend between these parameters and the quality of the
reconstruction. Therefore, these data are not precise enough to
determine  unambiguously the free-free and synchrotron parameters.
However, it is still important to look at all the available
information, since not all combinations of $\beta_{\rm ff}$ and
$\beta_{\rm syn}$ 
produce equally good CMB reconstructions.
In the case of the dust component (see bottom right panel of
Fig.~\ref{fig:errors_simu} and top panel of Fig.~\ref{fig:param_simu})
there is a visible correlation 
between the correct dust model and the case ordering. Therefore the
dust parameters can be determined by the method itself 
for the case of a spatially invariant dust model. In
a real data set, where spectral variability would occur, the picture
would not be so clear, but there would
still be some trends in the graphs that would point out to some
preferred models (see discussion in next section).  

We note that there are other possible
choices of weights that would assign the best $G$ to the
case with the correct spectral parameters. 
In fact, in practice, it would be very difficult to distinguish between
the five cases with a lower (and almost identical) value of $G$ for
this data set, 
since other similar choices of weights lead to a reordering of the top
cases. We remark, however, that the CMB reconstruction is very robust
independently of the chosen spectral parameters for those cases with 
low values of $G$. The fact that $G$ is so flat indicates again that
the data are too noisy to discriminate unambiguosly between the
different sets of spectral parameters. MEM is able to accomodate
a handful of different models within the noise, providing virtually 
indistinguishable CMB reconstructions.
\begin{figure}
\includegraphics[angle=-90, width=8cm] {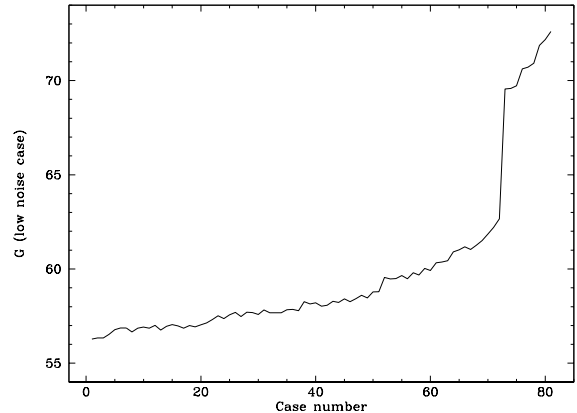}
\caption{Value of $G$ versus case number (ordered from low to high
value of $\chi^2$) for the low noise case.}
\label{fig:g_lownoise}
\end{figure}

It is interesting to look at the result if we apply the selection function $G$
to the low noise simulated case of the previous
section. Fig.~\ref{fig:g_lownoise} shows the values of the $G$
function versus the case number in the low noise case. Note that the
cases are ordered according to the value of the $\chi^2$ obtained
in the low noise reconstructions and therefore the numbering does not coincide
with the one on this section. As expected, the $G$ selector works also
in this case and the reconstruction with the
minimum value of $G$ coincides with the one obtained with the minimum
value of $\chi^2$ (which had the correct spectral
parameters). 
The fact that the value of $G$ is higher for the low noise case with
respect to the realistic case should not be surprising. $G$ can be
used to compare different reconstructions for the same
experiment. However it should not be used to
compare the quality of reconstructions for
experiments with different characteristics. For instance, MEM gives
more weight to minimising the value of the $\chi^2$ than that of the
entropy in the low noise case producing a higher value of S (although
with a slightly lower value of $\alpha$) what contributes to a higher
$G$. 
Also the value of $\sigma^{\rm rec}_{\rm CMB}$ is higher in the low noise case
since the CMB has been reconstructed with (intrinsic) higher resolution. 
A more interesting conclusion can be derived by looking at the range
of $G$ for the low and realistic noise cases. Whereas $G$ spans from
$\sim 56$ to $\sim 73$ for the low noise case, in the realistic case the
range for the same 81 cases goes from $\sim 54$ to $\sim 56$. This shows
again that the low signal to noise of the realistic data make it very
difficult to discriminate between the different considered cases.
We also note that the optimal choice of the 
coefficients defining $G$ may vary depending on the characteristics of the
considered experiment. Therefore, a detailed study using simulations
should be performed in each case.

\subsubsection{Reconstructed maps and power spectra for the
simulated case with realistic noise}

\begin{figure*}
  \begin{center}



    \includegraphics[width=16cm]{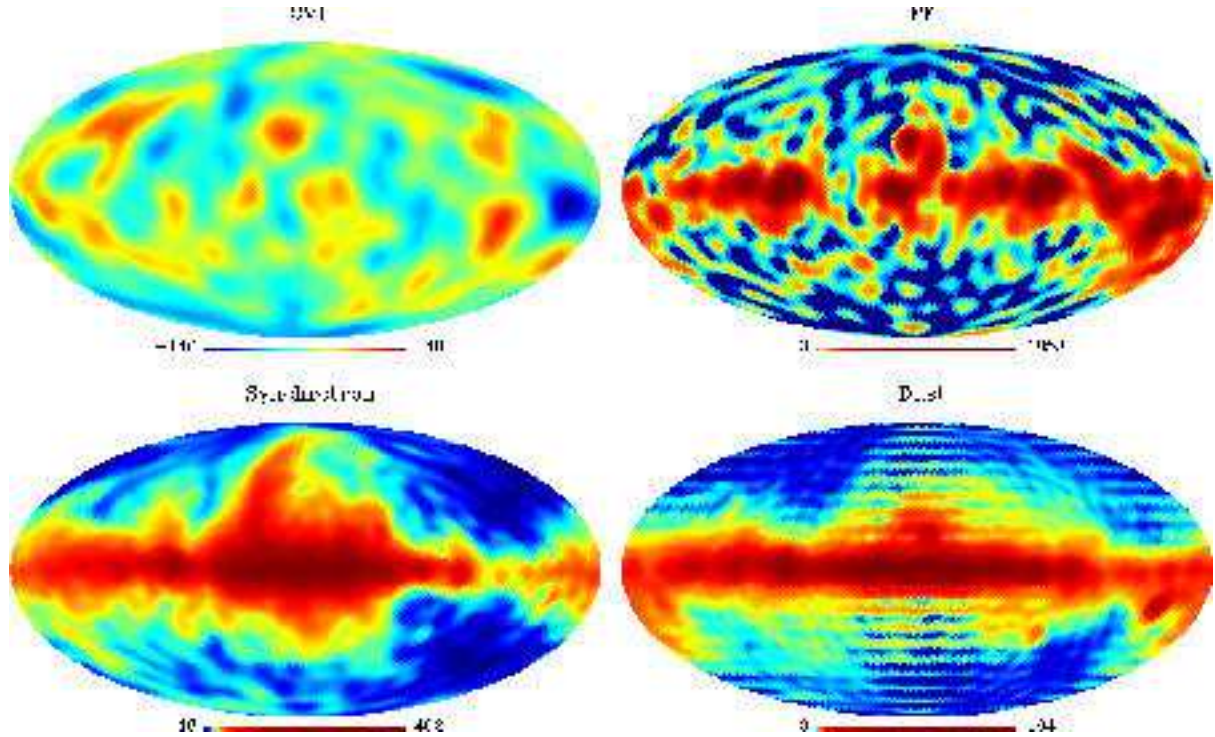}

  \end{center}
  \caption{\label{fig:rec_simu} Reconstructed components from
  realistic simulated data at 50~GHz obtained
  using the correct spectral parameters (case 1). The maps have been
  smoothed with a 7 degree Gaussian beam and have been plotted in the same
  scale as the input maps of Fig.~\ref{fig:components} to allow a
  straightforward comparison}
\end{figure*}

\begin{figure*}
  \begin{center}




    \includegraphics[width=16cm]{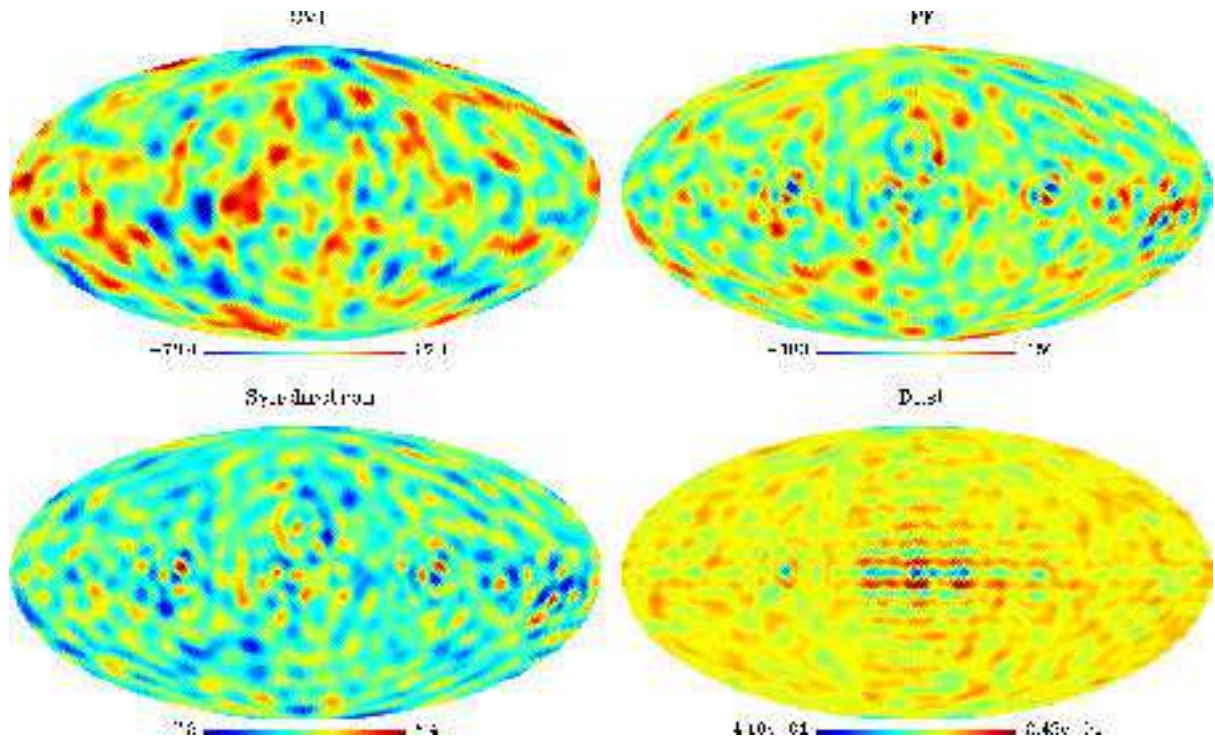}

  \end{center}
  \caption{\label{fig:residuals}Residuals of the simulated
  data for the correct spectral parameters (case 1) obtained by
  subtracting the smoothed input maps (Fig.~\ref{fig:components}) 
  from the smoothed reconstructions (Fig.~\ref{fig:rec_simu}).
  All the maps have been plotted in linear scale.}
\end{figure*}

We have plotted the reconstructed maps corresponding to case 1 
(i.e. using the correct spectral parameters) smoothed with a 7 degree
Gaussian beam in Fig.~\ref{fig:rec_simu} in the same scale as the
input maps (Fig.~\ref{fig:components}) to allow for a straightforward
comparison. The residuals map for each component is given in
Fig.~\ref{fig:residuals}. The values of the residuals dispersion for each
component (all-sky, high and low galactic latitude), the dispersion of the
input maps and the estimated errors are summarized in
Table~\ref{tab:rms_simu}.
\begin{table}
\begin{center}
\caption{Dispersion values for input, reconstructed (from case 1) 
and residuals maps smoothed with
a 7 degree beam at 50 GHz in $\mu$K. These values
are given for the whole map (col.~3) , for the region outside the
galactic cut (col.~4) and for the galactic centre (col.~5). 
For comparison the estimated error is also given in the last column.}
\label{tab:rms_simu}
\begin{tabular}{c|ccccc}
\hline
Cpt & Map & $\sigma_{all}$ & $\sigma_{out}$ & $\sigma_{Gal}$  & $e_{est}$\\
\hline
& Input & 36.4 & 36.5 & 36.3 & \\
CMB & Rec. & 34.3 & 33.2 & 37.4 & \\
& Resid. & 21.5 & 21.4 & 21.3 & 21.8 \\
\hline
& Input & 186.2 & 34.8 & 255.6 & \\
FF & Rec. & 187.2 & 46.9 & 256.3 & \\
& Resid. & 36.9 & 33.2 & 42.4 & 42.0 \\
\hline
& Input & 42.7 & 10.1 & 59.2 & \\
Synch. & Rec. & 42.8 & 10.2 & 59.4 & \\
& Resid. & 1.7 & 1.5 & 1.9 & 2.1 \\
\hline
& Input & 16.6 & 0.66 & 25.1 & \\
Dust & Rec. & 16.6 & 0.66 & 25.1 & \\
& Resid. & 0.04 & 0.03 & 0.05 & 1.3 \\
\hline
\end{tabular}
\end{center}
\end{table}

The dispersion of the CMB residuals on an angular scale of 7 degrees
is at the level of $\sim 21$ $\mu$K, 
which is in good agreement with the estimated error given by the
MEM algorithm (see Table~\ref{tab:rms_simu}). 
Many of the main features of the CMB input map are
also present in the reconstruction. However, the smallest structure
has clearly been damped in the reconstructed map. This is expected
since at the highest considered $\ell$'s the COBE-DMR 
data are dominated by noise and MEM just defaults to zero in absence of any
useful information.
It is interesting to point out that the CMB errors at high and low galactic
latitudes are actually comparable ($\sim 21 \mu$K in both cases,
see Table~\ref{tab:rms_simu}), showing that the map
is equally well recovered independently of the Galaxy. This is the
case even for small departures of the spectral parameters from the
true ones, such as those in the 5 best cases, which have very similar
residuals dispersions in both regions. This is due to the very
high noise of the COBE-DMR maps that allows to accommodate the data to
different spectral models. In fact, as already mentioned, the CMB
reconstruction is very robust for models with a low value of $G$. The
difference between the 7 degree smoothed CMB 
reconstruction of case 1 and those of cases 2 to 5 are $< 2\mu K$
outside the Galaxy and $ < 4\mu K$ inside the galactic
cut for all the cases, values which are significantly lower than 
the statiscal error. Therefore, even if we can not determine with total 
reliability the spectral parameters from our data set due to their low
signal-to-noise, we can avoid the introduction of errors in the CMB
reconstruction due to these uncertainties, especially at the high
galactic latitude region. For cases with larger values of $G$,
the reconstructed error inside the Galaxy starts to be systematically
higher than that of the high galactic latitude region.

Regarding the free-free map, it can be seen that the galactic plane is
reasonably well recovered (with an error $\sim$20 per cent) whereas
most of the structure outside the galactic cut has been lost. In fact
much of the signal recovered at high galactic latitude takes negative
values. 
The synchrotron and dust maps have been very well recovered since MEM
has succeeded in tracing these emissions from the Haslam and COBE-DIRBE
maps respectively. There are only some small differences between the
input and reconstructed maps, mostly at small scales, and the
dispersions of the residuals are at the level of $\sim 4$ per cent for
the synchrotron and $\sim 0.3$ per cent for the dust component. 

%

Of course, when the wrong spectral dependence is assumed, the errors
increase appreciably for all the galactic components.
However, these differences come mainly from a normalization factor
rather than from the spatial structure. 
For instance the reconstruction error for the
synchrotron in case 2 (where $\beta_{syn}=-0.9$ was assumed) is $\sim
17 \mu$K (versus $\sim 2\mu$K for case 1)
whereas the spatial cross correlation between the input and
reconstructed smoothed maps is at the same level (0.999) for both cases.
Thus, MEM is finding the right amplitude and structure of the 
synchrotron at the Haslam frequency, where this emission dominates, 
and then extrapolating to the reference frequency using the considered
spectral index. Similar ideas apply to the dust and free-free emissions.

\begin{figure*}
\includegraphics[angle=-90, width=16cm] {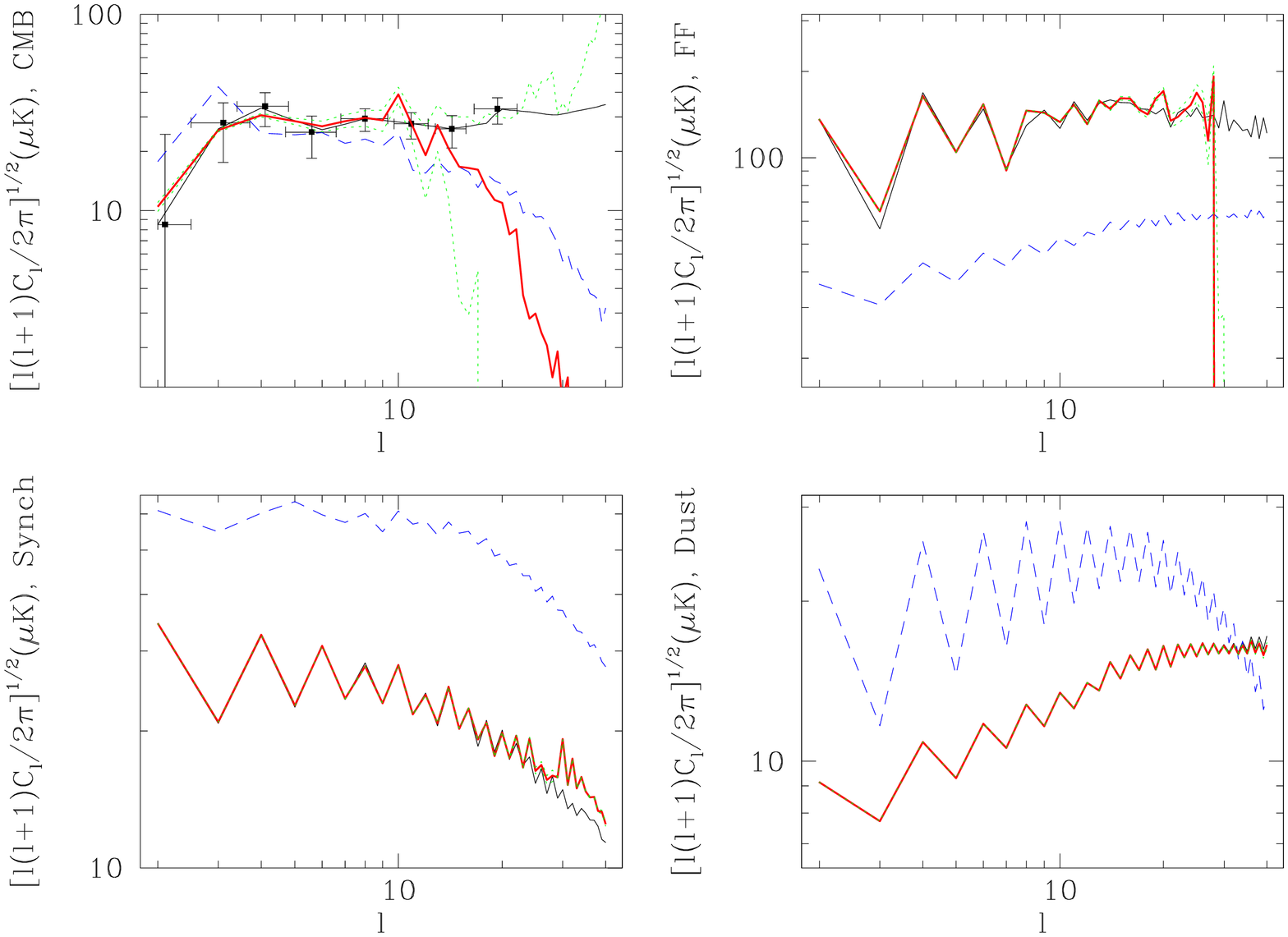}
\caption{Original (thin solid), initial guess (dashed) and 
reconstructed (thick solid) power spectra for the (realistic)
simulated data. The dotted lines indicate the estimated 1$\sigma$ confidence
level for the reconstructed power spectra. 
The correct spectral parameters have been used for the reconstructions
(case 1). For comparison the power spectra measurements obtained from
the COBE-DMR data are also shown (solid squares in top left panel,
taken from Tegmark \& Hamilton 1997).}
\label{fig:ps_simu}
\end{figure*}
\begin{figure*}
\includegraphics[angle=0, width=16cm] {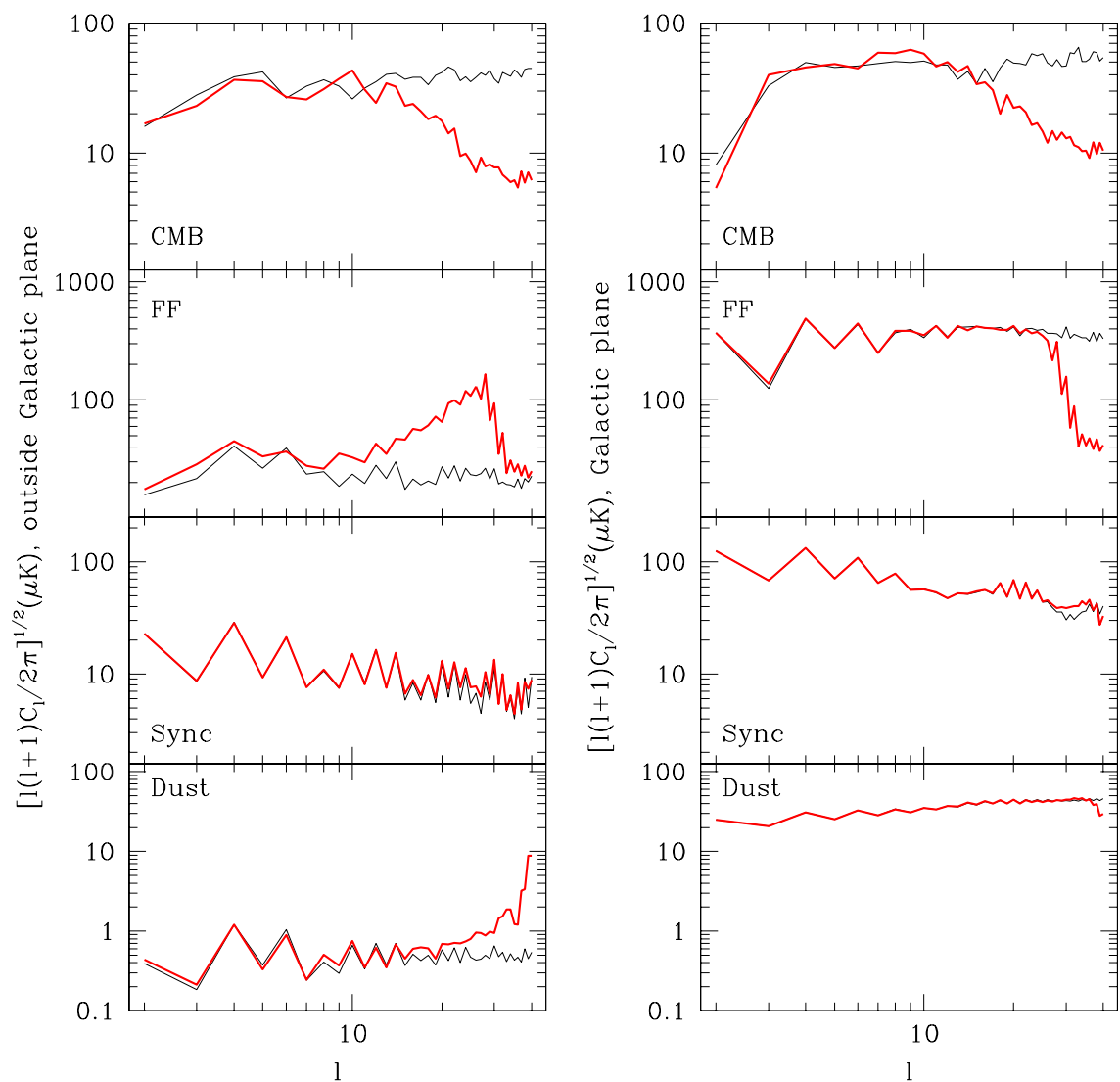}
\caption{Original (thin solid) and reconstructed (thick solid) power
spectra for the simulated data inside and outside the galactic plane.
The correct spectral parameters have been used for the reconstructions
(case 1).}
\label{fig:ps_simu_cut}
\end{figure*}

In Fig.~\ref{fig:ps_simu} the true (thin solid line) and reconstructed
(thick solid line) power 
spectra for the unsmoothed input and recovered maps have been plotted.
The dotted lines are the estimated
$1\sigma$ confidence level for the reconstructed components. Finally,
the dashed line correspond to the initial power spectra supplied to
the MEM algorithm.
In the CMB panel (top right) we have also plotted the power spectra
measured from COBE (solid squares).
Note that the CMB power spectrum used as initial guess in MEM
(dashed line) is
similar but differs from the true one (thin line),
allowing for possible errors in the estimation of the prior.
Even without iterating in the CMB power spectra, the recovered
$C_\ell$'s follow quite well the true ones up to $\ell \sim 15$ and then
start to drop due to the resolution of the COBE-DMR data.
As already mentioned, this reflects a loss of
resolution in the reconstructed map with respect to the input one.
We may also wonder about the quality of the recovered power
spectra inside and outside the galactic
cut. Fig.~\ref{fig:ps_simu_cut} shows the true and reconstructed power
spectra in these two regions of the sky. For simplicity,
we have estimated the $C_\ell$'s padding with zeros the masked region and they
have been rescaled according to the considered area. 
For the CMB case
the results are quite similar for both regions of the sky, and the
reconstructed power spectrum follows
approximately the true power up to $\ell \sim 15$, as it was the case
for the whole sky.

Regarding the galactic components, it is quite striking that MEM is
able to recover the right power spectra even when the initial guess
was far off from the correct $C_\ell$'s. The free-free power
spectra (top right panel of Fig.~\ref{fig:ps_simu}) has been
reasonably well recovered up to 
$\ell \sim 30$ (although some excess is present at the smallest
scales) where it sharply drops to zero. As in the case of the CMB, the
free-free information comes mainly from the noisy COBE-DMR data whose signal
to noise ratio and resolution does not allow MEM to 
provide a reconstruction at higher $\ell$'s.
An interesting result is also found by looking at
Fig.~\ref{fig:ps_simu_cut}. The reconstructed FF power spectrum inside
the galactic plane follows quite well the input one up to $\ell \sim
25$, however the reconstruction is much poorer outside the galactic
cut. In fact, a clear excess of power is seen at $\ell \ga 10$, which
gives rise to the spurious structure that is present in the FF
reconstructed map at high galactic latitude and to the excess of power
seen in the whole-sky reconstructed power spectrum. The reason for MEM
to fail in recovering the FF in this region of the sky is again the low
signal-to-noise ratio of the data together with the weakness of the FF
component outside the Galaxy.
The synchrotron is faithfully recovered up to $\ell \ga 20$,
result that holds for both the high and low galactic latitude regions
as well as for the whole map reconstruction. At higher
$\ell$'s there is some excess power which is responsible for the
structure seen in the synchrotron residuals.
Finally, the recovered dust power spectra in the galactic
region follows very well the input
one up to practically the considered $\ell_{max}$. This is due to the
fact that the dust emission is mainly recovered from the COBE-DIRBE
map, which provides the algorithm with enough resolution to recover faithfully
this component up to very high $\ell$. Regarding the high galactic
latitude region, there is an excess of power present at  $\ell \ga
20$. However this structure does not show in the global power spectrum
reconstruction, which is very good up to high $\ell$, since this is
dominated by the emission at the galactic region, which has been very
well recovered.

\section{Analysis of real data}
\label{sec:results_data}

We have applied our MEM algorithm to the set of data described in
$\S\ref{sec:data}$. As in the previous section, we have aimed
to reconstruct the CMB, free-free, synchrotron and dust emissions using
the COBE-DMR, Haslam and the lowest frequency COBE-DIRBE maps. 
As the initial guess for our reconstructions we have chosen the power
spectra of the simulated components of the previous sections (smoothed
with the COBE-DMR beam for the CMB and the
free-free\footnote[3]{The two all-sky templates for free-free emission
(Dickinson et al. 2003, Finkbeiner 2003)
became available shortly after the completion of this work.
Therefore, instead of our partially mock template, we could have used
 one of this new maps to compute 
 the initial free-free power spectrum.
 However, as shown using simulated data, the
 final result is very insensitive to the initial guess for the power
 spectrum. Thus we do not expect that our results would be
 appreciably modified by using a more realistic free-free template.}
 and with a Gaussian beam
of $\mbox{FWHM}=263.8$ arcmin 
for the dust and synchrotron). As before, we have iterated
over the power spectra for the galactic components but not for the
CMB, since we have some prior knowledge about its power spectra that
we would like to include in the analysis.

\subsection{Estimation of the spectral parameters}
\label{sec:freq_data}
\begin{figure}
\includegraphics[width=8cm] {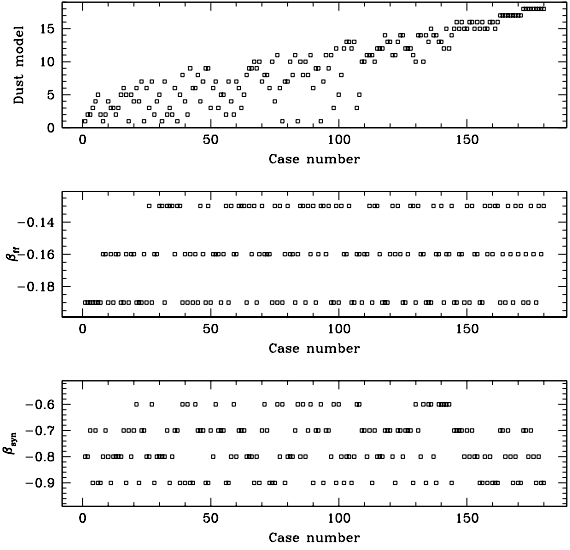}
\caption{Dust model (top), free-free spectral index (middle) and
synchrotron spectral 
index (bottom) versus the case number for real data. 
The dust model number correspond to: $T_d=19$,$\alpha_d=2$ (1),
$T_d=18$,$\alpha_d=2$ (2), $T_d=16$,$\alpha_d=2.2$ (3),$T_d=22$,$\alpha_d=1.8$
(4), $T_d=20$,$\alpha_d=2$ (5),$T_d=16$,$\alpha_d=2$
(6),$T_d=20$,$\alpha_d=1.8$ (7), 
$T_d=19$,$\alpha_d=1.8$ (8),$T_d=22$,$\alpha_d=2$
(9),$T_d=18$,$\alpha_d=2.2$ (10), 
$T_d=18$,$\alpha_d=1.8$ (11),$T_d=20$,$\alpha_d=2.2$
(12),$T_d=19$,$\alpha_d=2.2$ 
(13),$T_d=22$,$\alpha_d=1.6$ (14), $T_d=16$,$\alpha_d=1.8$
(15),$T_d=20$,$\alpha_d=1.6$ (16), 
$T_d=18$,$\alpha_d=1.6$ (17),$T_d=16$,$\alpha_d=1.6$ (18)}
\label{fig:param_data}
\end{figure}

\begin{figure*}
\includegraphics[width=16cm] {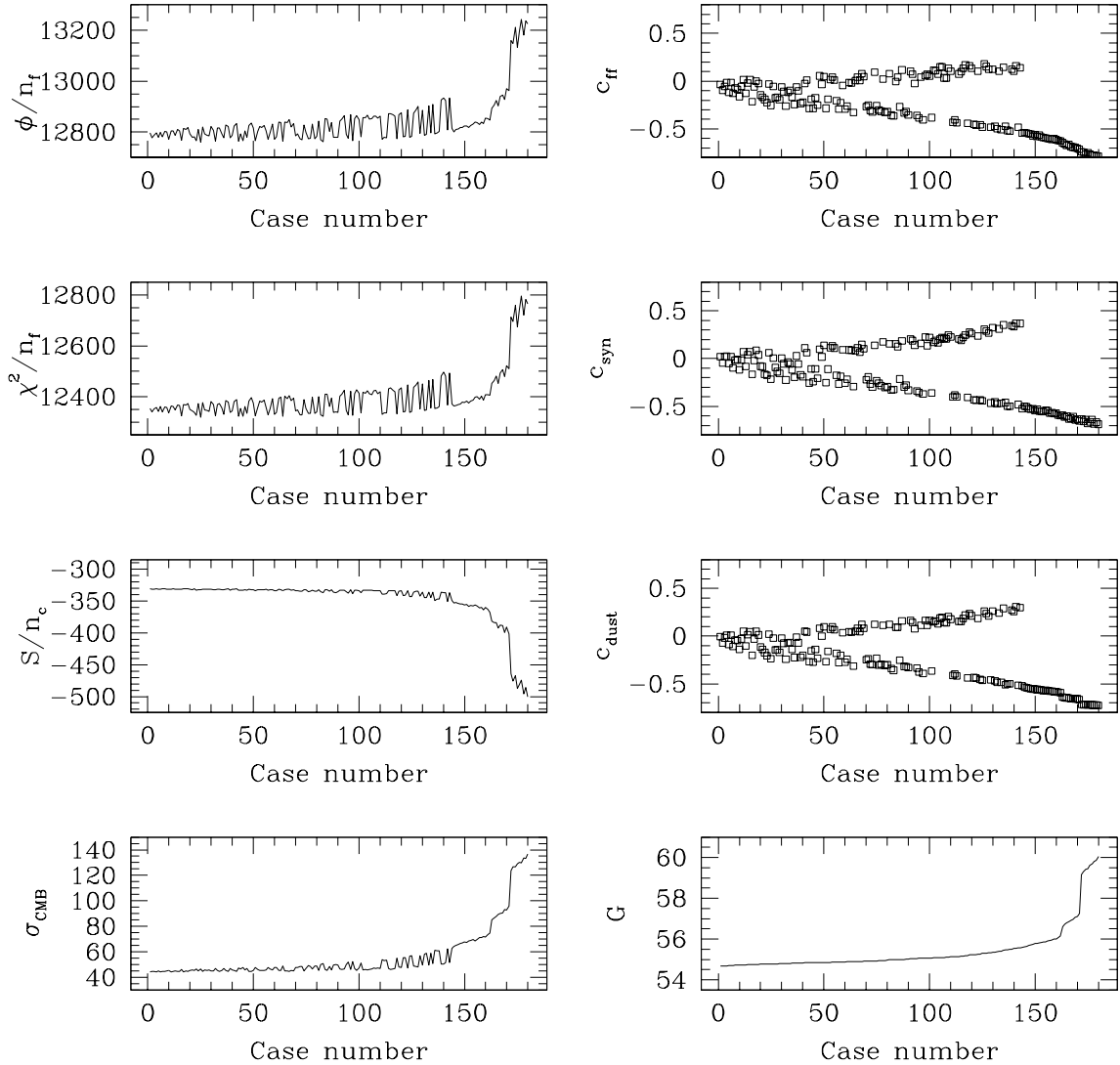}
\caption{As Fig.~\ref{fig:chi2_simu} but for the 180 cases used to recover the
microwave sky from the real data.}
\label{fig:chi2_data}
\end{figure*}

\begin{figure*}
\includegraphics[angle=-90, width=16cm] {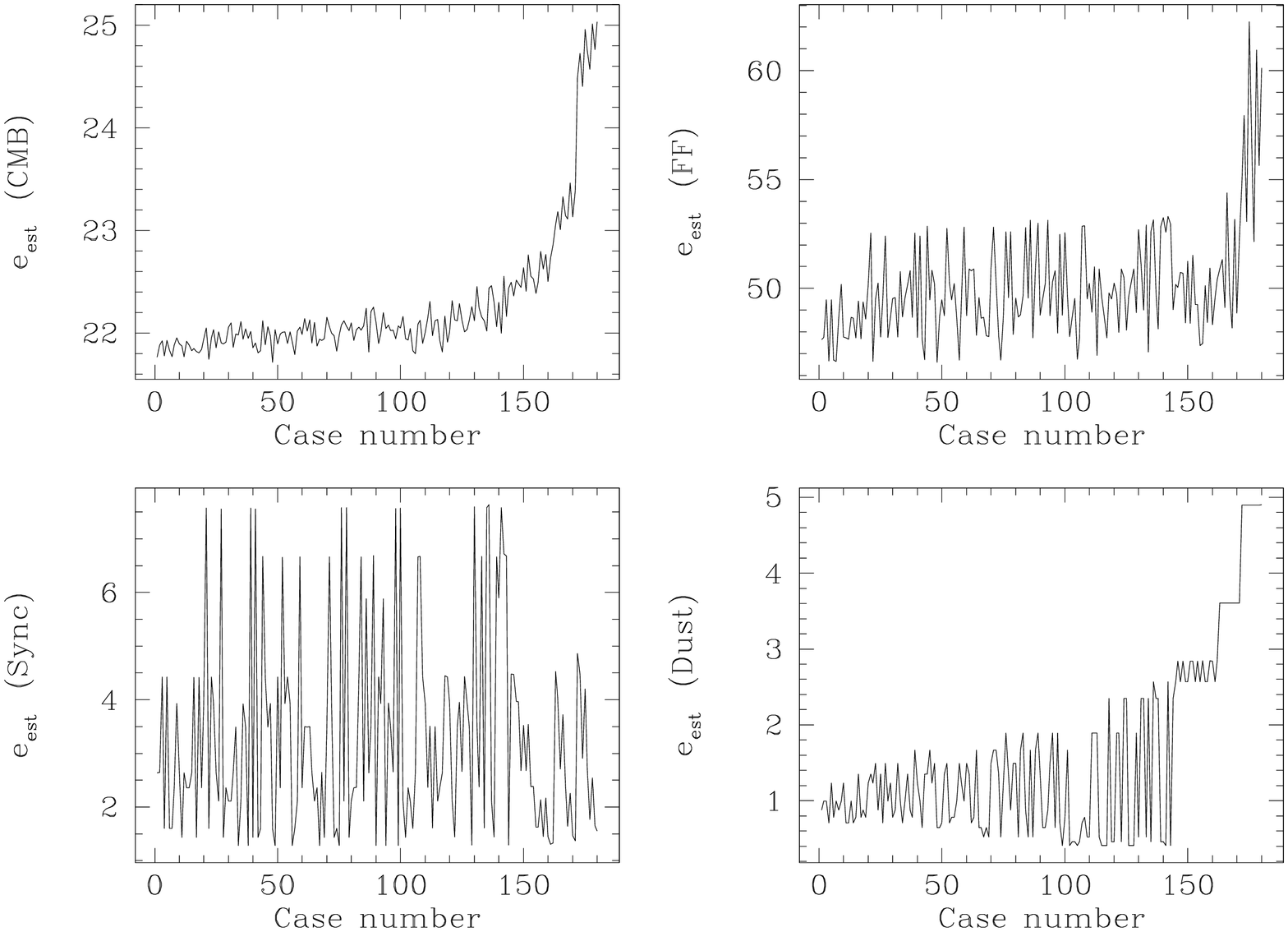}
\caption{Estimated errors of each of the reconstructed components
versus the case number for the real data. Each case corresponds to a
different set of spectral parameters and they have been ordered
according to the value of $G$.}
\label{fig:errors_data}
\end{figure*}

First of all we have applied MEM for 180 different sets of spectral
parameters with values in the range 16 to 22 K for the dust temperature,
1.6 to 2.2 for the dust emissivity, $-0.9$ to $-0.6$ for $\beta_{syn}$ and
$-0.19$ to $-0.13$ for $\beta_{ff}$. We have then ordered the different
cases according to the obtained value of $G$. The lowest value of $G$
is obtained for the parameters $T_d=19$ K, $\alpha_d=2$,
$\beta_{ff}=-0.19$ and $\beta_{syn}=-0.8$. 
Fig.~\ref{fig:param_data} shows the values of the spectral parameters
versus the case number.
In Fig.~\ref{fig:chi2_data} the value of $G$ as well as the quantities
used to calculate it are given for the 180 cases. It can be seen that
the trends of the curves are the same as those of the simulated cases
(compare with Fig.~\ref{fig:chi2_simu}). Not surprisingly, the value of 
$\varphi$, $\chi^2$, $|S|$ and $\sigma_{CMB}^{rec}$ 
are now a bit higher than in the simulated case, which indicates that our
model does not perfectly fit the data. This is due to the large number
of uncertainties present in the real case, such as the spectral
dependence of each component, its position and/or frequency
variability or even the number of components. However, the $\chi^2$ is
still very reasonable since it is of the order of the number of
data. 
As in the case of the simulated data, there is
a clear correlation between the dust model (top panel) and the value
of $G$, indicating that some of the studied dust parameters are
preferred by the data. In particular models with $\alpha_d=2$ and $T_d$
around 18-20 K are in the first positions. Conversely, the considered 
models with $\alpha_d=1.6$ are clearly disfavoured.
As expected, the distinction between the models is less clear than for
the simulated data, since degenerate cases are more important for true
data than for ideal ones.
Regarding the free-free spectral index (middle panel), there is a slight
trend from the data to favour those cases with more negative values of
$\beta_{ff}$. 
Values of $\beta_{syn}$ between $-0.9$ and $-0.7$ are basically
equally preferred by the data, whereas $\beta_{syn}=-0.6$ seems to
produce reconstructions with higher values of $G$.
Finally, for comparison with the ideal simulated case, we have plotted
the estimated error for the reconstructed components in
Fig.~\ref{fig:errors_data}. They are at similar levels as those in the
simulated cases.

Table~\ref{tab:best_data} gives the ten cases with the lowest
values of $G$. The difference between the CMB reconstruction for the 
selected case one and all the cases up to number 15
is between 0.8 and 2.8 $\mu$K outside the
galactic cut and between 1.6 and 10 $\mu$K in the galactic centre.
Therefore, as happened in the ideal case, the CMB recovered map in
the high galactic latitude region is very
robust against certain variations of spectral parameters.
\begin{table}
\begin{center}
\caption{Spectral parameters and value of $G$ for the ten best cases
from real data}
\label{tab:best_data}
\begin{tabular}{c|ccccc}
\hline
Case & $T_d$ & $\alpha_d$ & $\beta_{ff}$ & $\beta_{syn}$ & $G$ \\
\hline
1 & 19 & 2.0 & $-0.19$ & $-0.8$ & 54.6674 \\
2 & 18 & 2.0 & $-0.19$ & $-0.8$ & 54.6678 \\
3 & 18 & 2.0 & $-0.19$ & $-0.7$ & 54.6747 \\
4 & 16 & 2.2 & $-0.19$ & $-0.9$ & 54.6812 \\
5 & 22 & 1.8 & $-0.19$ & $-0.7$ & 54.6933 \\
6 & 20 & 2.0 & $-0.19$ & $-0.9$ & 54.6946 \\
7 & 18 & 2.0 & $-0.19$ & $-0.9$ & 54.7056 \\
8 & 19 & 2.0 & $-0.16$ & $-0.8$ & 54.7216 \\
9 & 18 & 2.0 & $-0.16$ & $-0.7$ & 54.7223 \\
10& 22 & 1.8 & $-0.19$ & $-0.8$ & 54.7227 \\
\hline
\end{tabular}
\end{center}
\end{table}

\subsection{Reconstructed maps and power spectra}
\begin{figure*}
  \begin{center}




    \includegraphics[width=16cm]{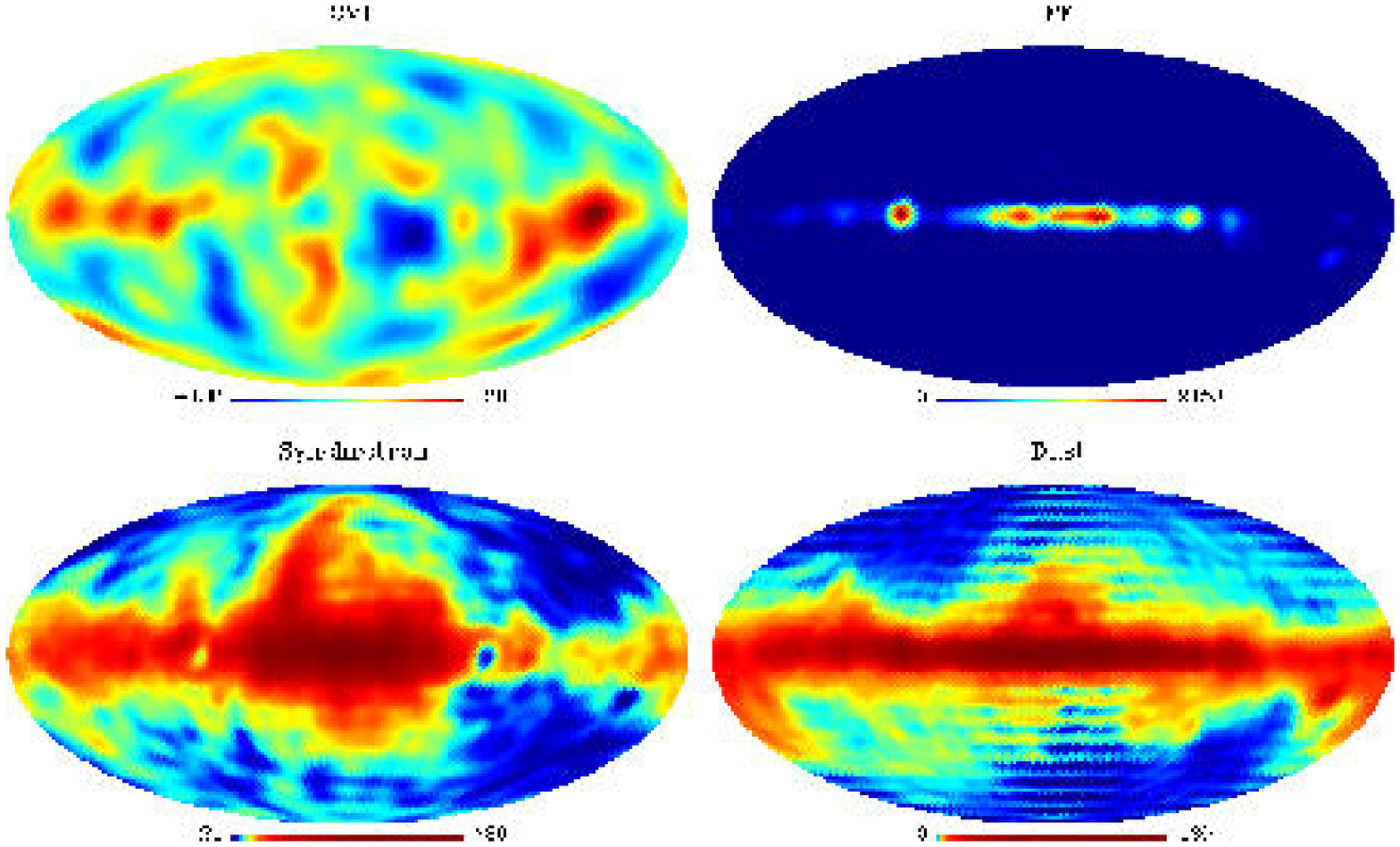}

  \end{center}
  \caption{\label{fig:rec_data}Recovered components smoothed with a 7
  degree Gaussian beam from real data 
  for the spectral parameters
  $T_d$=19~K, $\alpha=2$, $\beta_{ff}=-0.19$ and $\beta_{syn}=-0.8$
  (case 1) at 50 GHz in $\mu$K. The synchrotron and dust components
  have been plotted in a non-linear scale.}
\end{figure*}

Fig.~\ref{fig:rec_data} shows the reconstructed CMB, free-free, Synchrotron
and dust emissions for case 1 ($T_d=19$, $\alpha_d=2$,
$\beta_{ff}=-0.19$ and $\beta_{syn}=-0.8$). 
\begin{figure}
  \begin{center}
    \includegraphics[width=8cm]{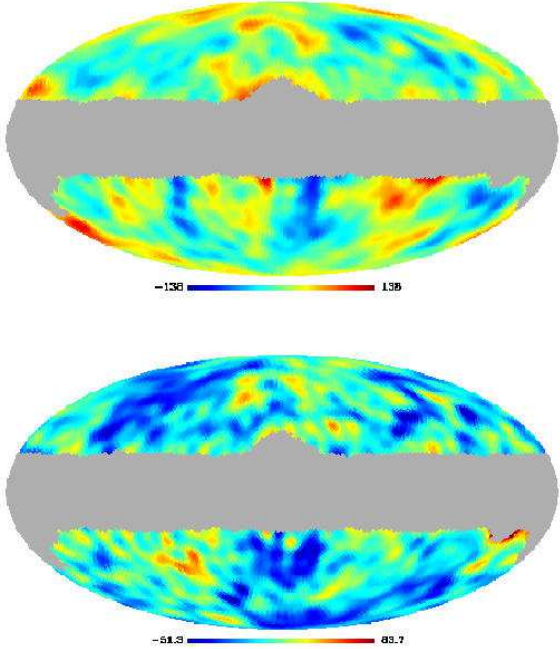}
  \end{center}
  \caption{\label{fig:53+90} Top: coadded 53 and 90 GHz COBE-DMR maps
  smoothed with a 7 degree Gaussian beam in
  $\mu$K outside the galactic cut. The map is plotted in the same scale as the CMB
  reconstruction of Fig.~\ref{fig:rec_data}. Bottom: the
  difference between the top map minus the CMB reconstructed map of
  Fig.~\ref{fig:rec_data} outside the galactic cut is shown. For the
  comparison, the monopole and dipole outside the galactic region have 
  been subtracted from the coadded and reconstructed maps.}
\end{figure}

\begin{table}
\begin{center}
\caption{Dispersion values for the reconstructed components from real data 
smoothed with a 7
degree Gaussian beam at 50 GHz in $\mu$K (for case 1) for the
realistic simulated data. These values
are given for the whole map (col.2), for the region outside the
galactic cut (col.3) and for the galactic centre (col.4). The
estimated error is also given in column 5.}
\label{tab:rms_data}
\begin{tabular}{c|cccc}
\hline
Component & $\sigma_{all}$ & $\sigma_{out}$ & $\sigma_{Gal}$ & $e_{est}$\\
\hline
CMB & 41.1 & 35.0 & 47.4 & 21.8 \\
FF & 238.4 & 40.9 & 360.7 & 47.7 \\
Synch. & 33.6 & 10.7 & 45.6 & 2.6 \\
Dust & 12.6 & 0.67 & 18.6 & 0.88 \\ 
\hline
\end{tabular}
\end{center}
\end{table}
Table~\ref{tab:rms_data} gives the dispersion level of the reconstructed
components smoothed with a 7 degree Gaussian beam 
at the reference frequency of 50 GHz, for the whole-sky, 
the Galaxy and the high galactic latitude region. For the CMB, we can
see that the dispersion for the two considered regions differ
appreciably (35 versus 47 $\mu$K). This indicates that there is
some contamination at low galactic latitudes in the CMB reconstructed
map. This is not surprising given the uncertainties
present in our model of the data (spatial variability of spectral
parameters, frequency model for each component, number of present
components, possible systematics, etc.).
The estimated error of the CMB reconstruction on a scale of 7 degree 
is $\sim 22 \mu$K. However, given the contamination of the CMB map in
the galactic 
centre, we can consider that this value is a fair estimate of the
CMB residuals only outside the galactic cut.

For comparison we have also plotted in Fig.~\ref{fig:53+90} a CMB map 
obtained by coadding the 53 and 90 GHz COBE-DMR maps
(each pixel weighted according to the inverse of its noise variance)
which has been smoothed with a 7 degree Gaussian beam. 
To allow for a better comparison, we have removed the monopole and
dipole outside the galactic cut which were mainly due to galactic
emission. As expected, we find this map to be well correlated with our
reconstructed CMB map (whose monopole and dipole have also been removed
outside the galactic mask to calculate this correlation), at the level
of 0.85 (outside the galactic cut), 
which confirms the presence of CMB signal in our reconstruction.
The difference between this coadded map and
our reconstruction outside the galactic cut is also shown in the same
figure (bottom). Note the smaller scale range in the difference map
which indicates the presence of common structure in both
maps. However, some clear structure is also seen in the difference
map. This is not surprising since not attempt to remove foreground
emission has been made in the coadded map.

Regarding the free-free reconstructed map (top right panel of
Fig.~\ref{fig:rec_data}), we are able to recover 
emission only inside the galactic cut. As in the simulated case, 
the free-free signal at high galactic latitude is lost. In fact, the
signal recovered outside the galactic cut takes mainly negative values
and its dispersion is lower than the estimated error ($\sim 48~\mu$K).
Regarding the galactic centre, MEM has only been able
to recover emission in $\sim 30$ per cent of the area of the galactic
cut, whereas the rest of the signal has zero or negative values.
The dispersion level of the 7 degree smoothed
free-free component in the galactic plane is $\sim 238 \mu$K considering all pixels
and it increases to $\sim 491 \mu$K if we consider only the fraction
of pixels with physical (i.e. positive) emission. This is several times
higher than expected from current $H\alpha$ estimations and
raises again the issue of an anomalous component. 
It is also interesting to note the high spatial
correlation between the galactic plane of the reconstructed free-free map and
that of the COBE-DMR frequency channels.

The synchrotron reconstructed map (bottom left panel of
Fig.~\ref{fig:rec_data}) presents structure both inside and outside
the Galaxy. As expected, the reconstructed emission is very well
correlated with the Haslam map (0.96 for the smoothed maps). 
We find a level of $\sim 46 \mu$K for the synchrotron emission in the
Galaxy and of $\sim 11\mu$K outside the Galaxy. The estimated error in
the reconstruction is $2.6 \mu$K. We should note, however, that this
is the statistical error associated to our method but it does not take
into account the uncertainties in the determination of
$\beta_{\rm syn}$.

Finally, the bottom right panel of Fig.~\ref{fig:rec_data} shows the
reconstructed dust emission smoothed with a 7 degree Gaussian beam. 
The visible ringing is an artifact due to the fact that we
are recovering the dust only up to $\ell_{max}=40$, whereas the DIRBE
data map (from which the algorithm traces this component) has power at
higher multipoles. We have found a dispersion value for the dust component 
of $\sim 19 \mu$K and of
$\sim 0.7 \mu$K inside and outside the Galaxy respectively at 50 GHz.
The (statistical) estimated error is below 1~$\mu$K.

\begin{figure*}
\includegraphics[angle=-90, width=16cm] {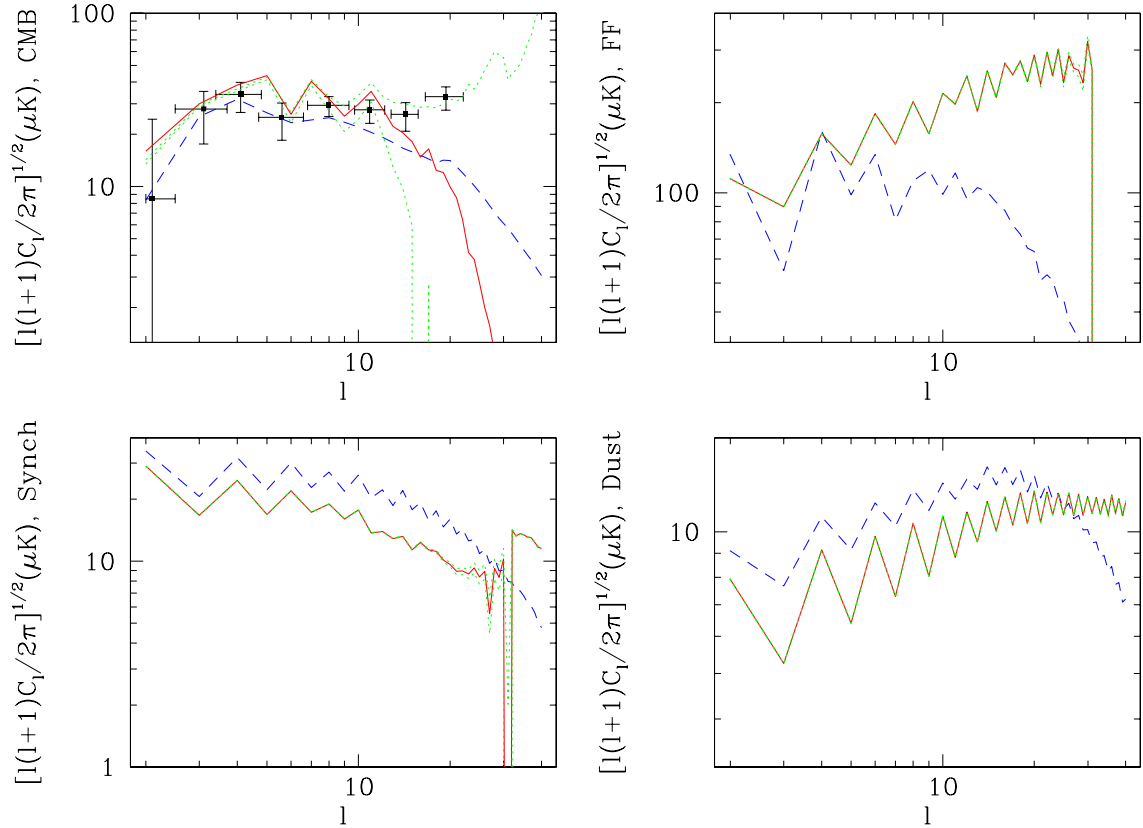}
\caption{Initial guess (dashed line) and reconstructed (solid line)
power spectra for each of reconstructed components obtained from the
real data for case 1 at 50 GHz. The dotted lines correspond to the $1\sigma$
confidence level of the reconstructed power spectra. For comparison
the power spectrum measurements obtained from the COBE-DMR data are
also plotted in the CMB panel.}
\label{fig:ps_data}
\end{figure*}

\begin{figure*}
\includegraphics[angle=0, width=16cm] {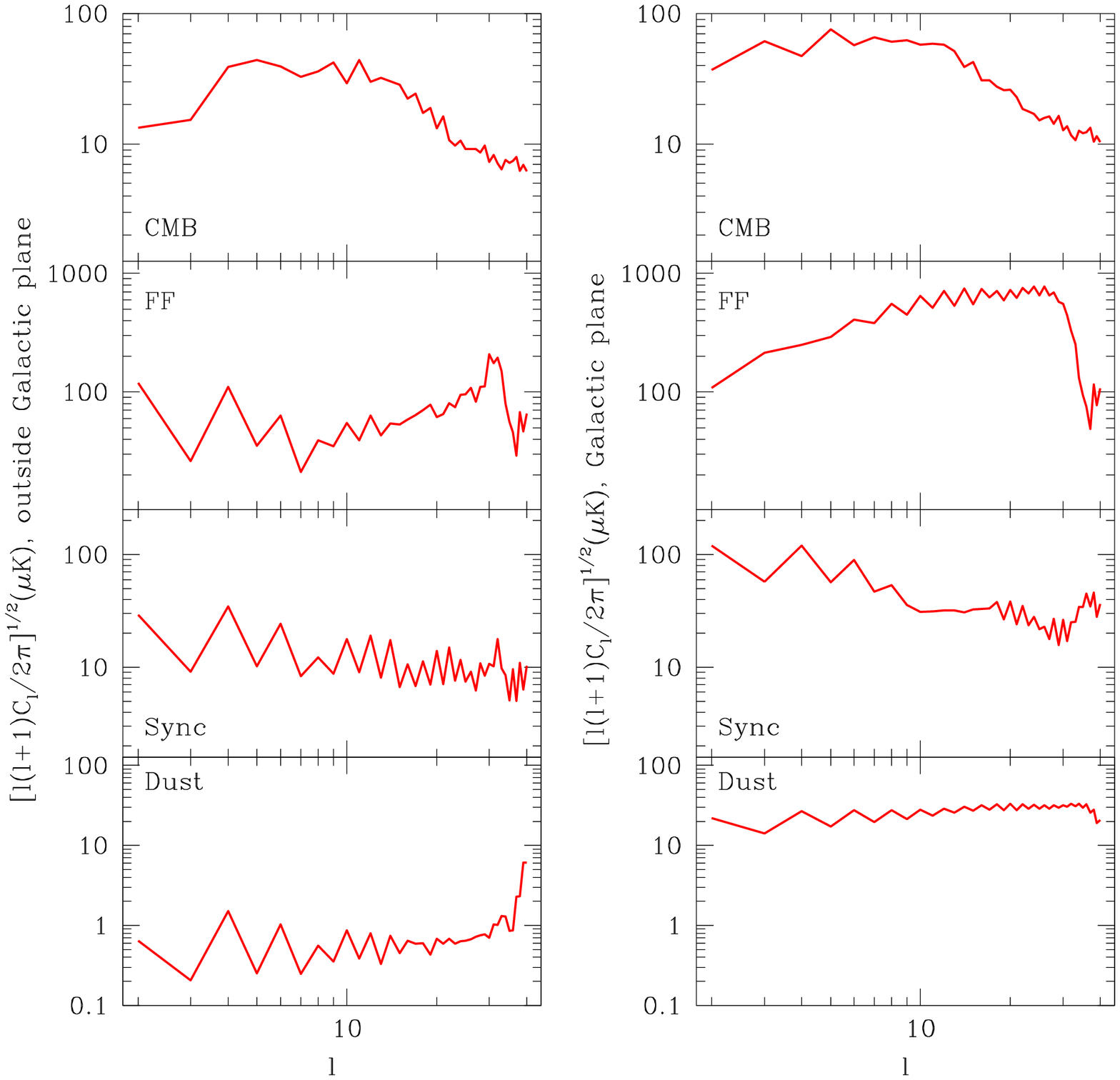}
\caption{Reconstructed power spectra for case 1 inside and outside the
galactic cut for real data.}
\label{fig:ps_data_cut}
\end{figure*}

The (unsmoothed) reconstructed power spectra (solid line) for the four
recovered components are given in Fig.~\ref{fig:ps_data}. The dashed
lines show our initial guess for the power spectra and the dotted
lines indicate the 1$\sigma$ confidence level for the reconstructed
power spectra. Fig.~\ref{fig:ps_data_cut} shows the power spectra for
the four reconstructed components inside and outside the galactic
cut, that have been estimated in the same way as for the simulations.

For comparison we have also plotted the CMB measurements obtained for
COBE-DMR in the CMB panel (top-left) of Fig.~\ref{fig:ps_data}. It is
interesting to 
note that the recovered $C_\ell$'s follow the shape of the power
spectra measured from the COBE-DMR data but with a higher
normalization. In addition, the reconstructed CMB power spectra
for the high galactic latitude region (top left panel of
Fig.~\ref{fig:ps_data_cut}) has approximately the expected
amplitude whereas the one at the Galaxy (top right panel of
Fig.~\ref{fig:ps_data_cut}) presents an excess of power at
all scales. These are also indications of the fact that we have
some galactic contamination in our reconstructed map, especially at
high galactic latitudes.

The free-free reconstructed power spectra presents the same behaviour as in
the simulated case. It seems to be recovered up to $\ell = 30$ and then
drops sharply to zero, due to the low resolution of the COBE-DMR
channels, which MEM has mainly used to recover the free-free signal. 
We also find in the reconstructed power spectra inside and
outside the Galaxy the same pattern as in the simulations. In particular,
it seems to be an excess of power at large $\ell$ at high galactic
latitude, which
produces all the spurious signal that is found in this region of the
FF reconstructed map.

The shape of the synchrotron reconstructed power spectra (bottom left
panel) follows quite
well that of the input one up to $\ell \sim 25$ but with a lower
normalization. This is expected since the initial power spectra has
been obtained from a smoothed extrapolation of the Haslam map 
up to the reference frequency using a spectral
index of $-0.8$ (the same value as $\beta_{\rm syn}$ in our case). 
This extrapolation provides therefore an
upper limit for the reconstructed synchrotron emission 
(since the Haslam map is dominated by this component but also contains
contributions from other emissions), which is consistent with our result.
At higher $\ell$'s, the power spectrum starts to oscillate wildly due
to the lack of information in the data to recover this emission.

The bottom right panel shows the reconstructed $C_\ell$'s for the
dust component. As in the case of the simulations, the
power spectra seems to be well recovered up to $\sim \ell_{max}$.
As expected it follows the shape of the initial guess (the
differences at the higher $\ell$'s are due to the fact that the input has
been convolved with a Gaussian beam of FWHM=263.8 arcminutes)
which has been obtained by interpolating the COBE-DIRBE channel down
to 50 GHz using a dust model with $T_d=18$ and $\alpha_d=2$ (different
from the best model found for the reconstructions). As in the
simulated case, we also find what seems to be an excess of power at 
high $\ell$ values for the reconstructed power spectrum outside the 
galactic plane.

\section{Discussion and conclusions}
\label{sec:discussion}

We have presented a flexible MEM algorithm that combines the
advantages of real (or data) space and harmonic space. On the one hand, 
the $\chi^2$ is calculated in data space, which allows one to include
straightforwardly the properties of the noise as well as incomplete sky
coverage. On the other hand, the entropy is estimated in harmonic
space allowing easy introduction of available prior information about the
power spectra of the components that we aim to reconstruct.
In addition, the method takes into account correlations between
different modes because we perform a single minimisation of $\sim n_c
\ell_{max}^2$ variables instead of minimising mode-by-mode as in
harmonic MEM.
We perform this global minimisation using a Newton-Raphson method,
which needs the Hessian matrix ${\mathbf H}$ to be evaluated and inverted. In
practice, this is not feasible since we have a very large number of
variables so we have approximated ${\mathbf H}$ by a block
diagonal matrix, which gives very good results.
Unfortunately, this minimisation is very time consuming and the method
is many times slower than harmonic MEM. 

To test the performance of the method we have applied it to simulated
spherical data and then to real data. In particular, we have used the
three frequency channels of COBE-DMR, the Haslam map and the lowest
frequency map of COBE-DIRBE to reconstruct the CMB, free-free, synchrotron
and dust emissions.

\subsection{Analysis of simulated data}
An important issue that has been thoroughly studied in the present
work is the determination of the frequency dependence of the different
components. We have modelled the free-free and synchrotron with a power law
parametrised by indices $\beta_{\rm ff}$ and $\beta_{\rm syn}$ respectively
and the dust by a grey body model with two parameters (the dust
temperature $T_d$ and the dust emissivity $\alpha_d$). We have then
applied the MEM algorithm to our simulated data (generated using
$\beta_{\rm ff}=-0.16, \beta_{\rm syn}=-0.8, {\rm T}_d=18$ and $\alpha_d=2$) 
for many combinations 
of the spectral parameters and have studied the behaviour of different
diagnostic quantities when the assumed values of these 
parameters depart from the true ones.

Firstly, we have studied a low noise case where the noise level
of the COBE-DMR simulated maps have been lowered by a factor of 5. In
this case we find that the $\chi^2$ value of the reconstructions can
successfully identify 
the correct set of spectral parameters. In addition,
we also find that there exists a high correlation between the error of
the CMB reconstruction and other quantities that can be directly estimated 
from the reconstructed maps.
In particular, together with the $\chi^2$ value, the entropy, the
$\varphi$ function , the dispersion of 
the CMB reconstructed map, and the cross correlations between the
reconstructed CMB and galactic components are good indicators of the
quality of the reconstructions. 

\begin{figure}
\includegraphics[angle=-90, width=\hsize] {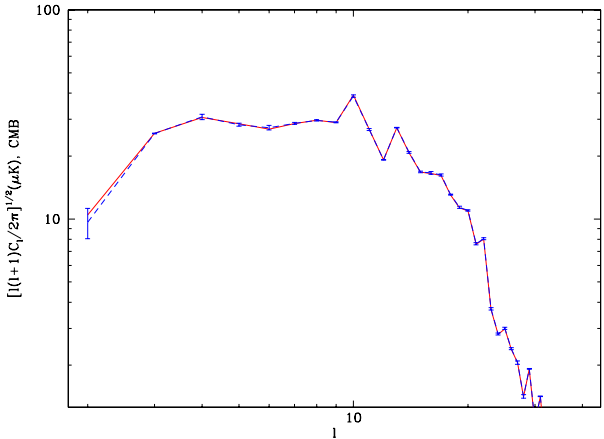}
\caption{CMB reconstructed power spectrum for case 1 of the simulated
data (solid line) and average CMB reconstructed
power spectra for the 5 cases with a lower value of $G$ (dashed
line). The error bars are the dispersion obtained from these 5 power
spectra. }
\label{fig:psmean_simu}
\end{figure}

Secondly, we have applied MEM to our set of simulated data using
realistic levels of noise for the same sets of spectral parameters. 
Due to the low signal to noise ratio of our data set, we find that the
$\chi^2$ is not longer able to pick the right combination of spectral
parameters. Therefore we have constructed an empirical selection function $G$
which is given by a linear combination of $\chi^2$, $|S|$, 
$\varphi$ , the dispersion of 
the CMB reconstructed map, and the cross correlations between the
reconstructed CMB and galactic components.
We pick as our best set of reconstructions the one with the lowest
value of $G$. We have considerd a total of 81 different cases, with 
three possible values for each spectral parameter. 
Using the quantity $G$, we can clearly see that the correct dust model is
clearly preferred over the others (see
Fig.~\ref{fig:param_simu}). However, it is very difficult to pick the
correct values of the spectral indices for the free-free and synchrotron,
since our data do not have enough information. In fact the
quantity $G$ is almost identical for the 5 best cases due to the fact
that our noisy data can accomodate a range of different spectral models.
In any case, even if we can not determine unambiguosly the spectral
parameters for the free-free and synchrotron component, the CMB
reconstruction is very robust outside the galactic cut and the 
differences between the CMB reconstructions for the cases with 
lower values of $G$ are well within the statistical errors.
In particular the level of the dispersion of the residuals of the smoothed 7
degree reconstructed CMB map outside the galactic cut 
ranges between 21.2 and 22.0 $\mu$K for all the cases between 1 and 30.
To illustrate this point we can also look at the differences in
the CMB reconstructed power spectra for the different cases. 
Fig.~\ref{fig:psmean_simu} shows the
CMB power spectrum obtained for case 1 (solid line) versus the
average power spectrum obtained from the 5 cases with lower values of
$G$. The error bars correspond to the dispersion in the values of the
$C_\ell$'s obtained from the former 5 cases. Note that this dispersion
is very small, which shows that the recovered CMB power
spectrum is also very robust for these 5 cases. This also indicates
that, at least for ideal simulated foregrouns with spatially constant
spectral parameters, the error introduced in the CMB power spectrum
due to the wrong identification of these parameters (provided
we pick one of the reconstructions with a lower value of $G$) is much
smaller than the statistical error.

For case 1 (the case with the correct spectral parameters) we have
found that the CMB error reconstruction for the (realistic) simulated
data is at 
the level of 21 $\mu$K. The level of CMB residuals are comparable
inside and outside the galactic cut, which indicates that the CMB has
been equally well recovered in both regions of the sky. Note that
the estimated error given by the method is $\sim 22 \mu K$.
The recovered CMB power spectrum follows the input one up to $\ell \sim
15$ and then starts to drop due to the resolution of the COBE-DMR
data. This result is again comparable inside and outside the
galactic plane. The free-free emission has been recovered with a
residual error of $\sim 42 
\mu K$ in the galactic plane at 50 GHz. At high galactic latitude, the
free-free 
reconstruction has basicly been lost. This result also shows in the
reconstructed power spectrum, where we find a clear excess of
structure at $\ell \ga 10$ outside the galactic cut, whereas MEM is
able to recover it up to $\ell \sim 25$ inside the Galaxy.
The synchrotron and dust emissions have been very well recovered
in the galactic plane with 
errors of $\sim 3$ per cent and $\sim 0.2$ per cent,
respectively. Outside the galactic cut the reconstruction is also
quite good, at the level of $\sim 15$ per cent for the synchrotron and
$\sim 5$ per cent for the dust. 
The whole-sky power spectra follows also quite faithfully the
input one up to $\ell \la 30$ for the synchrotron and up to
$\ell_{max}$ for the dust. However, the recovered dust power
spectrum shows some excess at small scales when only the region
outside the galactic cut is considered.
Note that MEM has been able to recover the power spectra of the
galactic components, independently of the supplied initial power
spectra (dashed lines in Fig.~\ref{fig:ps_simu}), which
was very far from the true one.
We have tested that different initial power spectra lead to very
similar reconstructions, provided MEM uses the
iterative mode to obtain the correct power spectra.

\subsection{Analysis of real data}
In $\S$\ref{sec:results_data} we have applied the same method to the
true data. We have calculated the value of $G$ for a total of 180
different sets of spectral parameters in the range $ -0.19 \le
\beta_{\rm ff} \le -0.13$, $-0.9 \le \beta_{\rm syn} \le -0.6 $, $16 \le T_d
\le 22 $ and $1.6 \le \alpha_d \le 2.2 $. The best reconstructions are
found for $\beta_{\rm ff}=-0.19$, $\beta_{\rm syn}=-0.8$, $T_d=19$ and
$\alpha_d=2$. The cross correlations between the CMB and galactic
components are at a similar level to those of the simulations
(Fig.~\ref{fig:chi2_data} versus Fig.~\ref{fig:chi2_simu}). However,
the value of $\varphi$, $\chi^2$, $|S|$, $\sigma_{CMB}^{rec}$ and thus of
$G$ are a bit higher than the ones found for the simulated cases.
This indicates that our model does not fit the data as well as in the
ideal case. This is not surprising taking into account the large
number of uncertainties of the data: the frequency dependence of the
model, variation of the spectral parameters across the sky or in the
frequency range or even the presence of some unknown component.
In spite of this, we still have a reasonable fit to our data (the
reduced $\chi^2$ is $\sim$ 1). As in the simulated case, some dust
models are clearly favoured with respect to the others
(Fig.~\ref{fig:param_data}) although the situation is not so clear as in
the ideal case. Again this is due to all the uncertainties we have
with respect to the frequency behaviour of the components which make
the presence of degenerate cases even more important than in the
simulated case.

\begin{figure}
\includegraphics[angle=-90, width=\hsize] {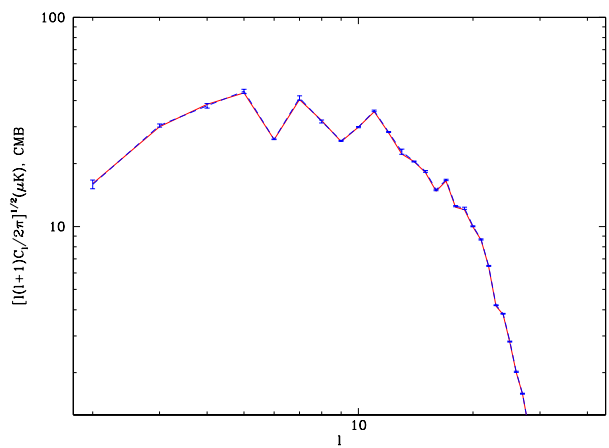}
\caption{CMB reconstructed power spectrum for case 1 of the real
data (solid line) and average CMB reconstructed
power spectra for the 5 cases with lower values of $G$ (dashed
line). The error bars are the dispersion obtained from these 5 power
spectra. }
\label{fig:psmean_data}
\end{figure}

The reconstructed maps for the four components are given in
Fig.~\ref{fig:rec_data} 
for case 1 (the one with the lowest values of $G$) and the
corresponding dispersions of the reconstructions as well as the
estimated errors are shown in Table~\ref{tab:rms_data}. From these
results, we can see that there is some clear contamination of the CMB
reconstruction in the galactic plane. For instance the dispersion of
the recovered CMB is at the level of 35 $\mu$K outside the galactic
cut, whereas in the galactic centre has a value of $\sim 47
\mu$K. This is another indication of the fact that our model for the
galactic components is not properly fitting the data. However, we believe
that the reconstructed CMB sky outside the Galaxy is reasonably well
recovered. Even if our assumptions about the spectral behaviour of the
components are wrong, we have seen in the simulations that 
this had very little effect in the CMB reconstructed map outside the
galactic cut, providing we are taking cases with low values of $G$.
This is also happening when using real data. 
In particular, the difference
between the 7 degree smoothed CMB restored maps at high galactic latitude for 
case 1 and all cases up to case 20 is $\la 3 \mu$K.
For the same cases, the differences are $\la 10 \mu$K in the galactic
plane, except for one of the cases.
We have also plotted the average and the dispersion of the CMB
recovered power spectrum for the 5 cases with a lower value of $G$
(Fig.~\ref{fig:psmean_data}). As in the simulated case, we find very little
dispersion within these reconstructed power spectra, which
confirms that the CMB reconstruction is not affected by certain
variations on the spectral parameters.

As in the simulated case, the free-free emission has been recovered only in
the galactic centre with a dispersion value at 50 GHz of $\sim 238 \mu$K 
and with an estimated error of $\sim 48 \mu$K. This value is several
times higher than expected from $H\alpha$ measurements and raise again
the issue of an unknown component.
The synchrotron and dust reconstructions, however, look very good with
estimated errors of $\sim 3$ and $\sim 1\mu$K respectively. These
values account only for the statistical error but do not
take into account the uncertainties coming from the determination of
the spectral parameters.

As shown in Fig.~\ref{fig:ps_data} the CMB recovered power spectrum
follows the shape of the measurements 
obtained from the COBE data, but with a slightly higher normalisation
which indicates again the presence of some excess contamination in the
recovered CMB map.  This is confirmed by looking at the
reconstructed CMB power spectrum inside and outside the galactic cut
(Fig.~\ref{fig:ps_data_cut}). At the high galactic latitude region the
amplitude of the $C_{\ell}$'s is consistent with the expected level
whereas those obtained for the galactic plane present an excess of power. 
The free-free is recovered up to $\ell \la 30$ and then drops to zero,
since the data do not have enough information at those
scales. As in the simulated case, what seems a spurious excess of 
power is found at high $\ell$'s in the reconstructed power spectrum of
the high galactic latitude region.

The recovered synchrotron power spectrum follows approximately the
shape of the input power 
spectra (as expected since it has been obtained extrapolating the
Haslam map) up to $\ell \la 25$ . For higher multipoles, the power 
spectra starts to oscillate which indicates that the reconstruction is
not reliable at those scales. 
Finally, the dust emission seems to be well reconstructed up to the
considered $\ell_{max}$ for the whole sky power spectrum. However,
some excess is found at high $\ell$'s in the power spectrum obtained
from the high galactic latitude region. We would like to point out
that, as in the simulated case, changing the initial guess power
spectra has very little effect in the reconstructions.

\begin{figure}
  \begin{center}



    \includegraphics[width=8cm]{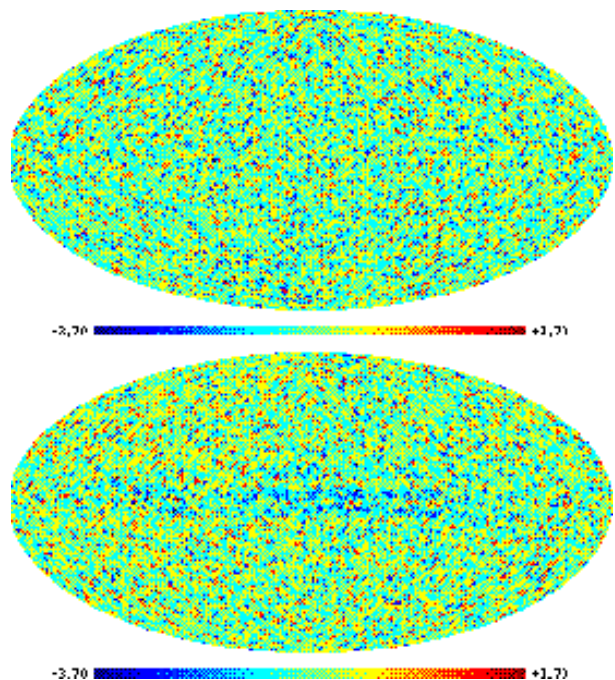}

  \end{center}
  \caption{\label{fig:resnoise02} Normalised residual noise for the
  COBE-DMR 53 GHz channel for case 1 of simulated data (top) and of real
  data (bottom).}
\end{figure}
An extra check of the quality of the reconstructions can be performed by
looking at the normalised residual noise for each frequency channel.
This map is constructed by subtracting the
predicted noiseless data (obtained 
using the reconstructed maps) from the actual noisy data. This gives
the `predicted noise' at the given frequency channel, which is then
divided by the noise dispersion at each pixel. If the fit is
acceptable this map should be a realisation of a
Gaussian white noise field of unit dispersion with no visible
structure, although one should be careful since incorrect choices of the
spectral parameters can also give rise to this result due to
degeneracies.
As an illustration, the top panel of Fig.~\ref{fig:resnoise02} shows the
normalised residual noise for the 53~GHz frequency channel for case 1
of the simulations. 
No visible structure is present in the map, which confirms on
the one hand, that the anisotropic noise has been properly taken into
account and, on the other hand, that the Galaxy has been well fitted
by our components. The bottom panel of the figure shows the same map
for case 1 of the real data, which shows some visible structure in the
galactic centre. This further
confirms the fact that we are not fully subtracting the Galaxy.

\subsection{Two-components dust model}
This inability of the components to fit the data may be due 
either to variations of the spectral parameters with
position or the frequency range, or to the fact that some other unknown
component is needed to fit properly the data.
A point raised by Jones et al. (1999) is the possiblity of reconstructing
more than one dust component to account for spatially variations of
the dust spectral parameters. Although each of the dust reconstructions may
not have physical meaning, adding them together improved the
reconstruction.
Moreover, when a dust component with emissivity variations from pixel 
to pixel was included in the simulated data, the errors of the CMB
reconstruction were also reduced by using three dust components with
different emissivities, as compared with the case of reconstructing 
a single dust component whose emissivity was the average of the input
template.

\begin{figure}
\includegraphics[angle=-90, width=\hsize] {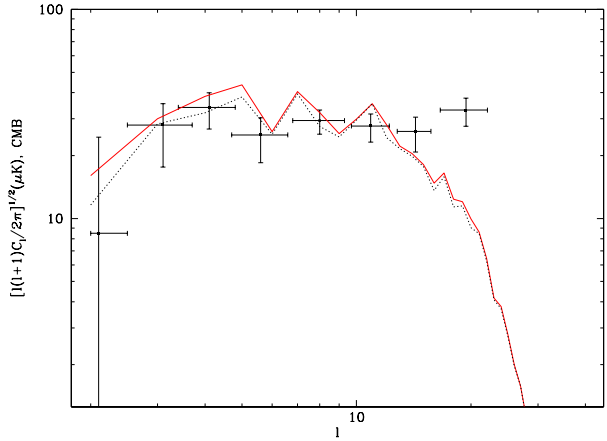}
\caption{Power spectrum obtained from the (unsmoothed) CMB
reconstruction of Fig.~\ref{fig:rec_data} (solid line) versus the
power spectra of 
the recovered CMB obtained using two dust components (dotted line).
The power spectrum measurements obtained from the COBE-DMR data are
also plotted (solid squares).}
\label{fig:ps_data_cmb}
\end{figure}

\begin{figure}
  \begin{center}



    \includegraphics[width=8cm]{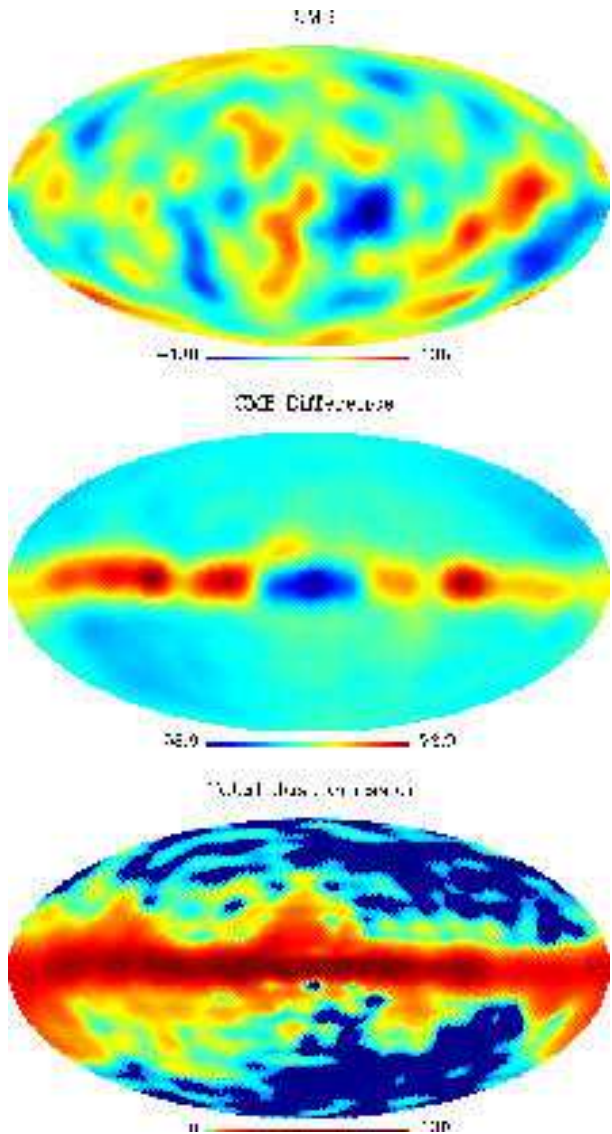}
  \end{center}
  \caption{\label{fig:2dust} Top panel: CMB reconstructed map at 50 GHz
  smoothed with a 7 degree Gaussian beam for the case when 2 dust
  components have been used. Middle panel: map resulting of
  subtracting the previous reconstruction from the one of
  Fig.~\ref{fig:rec_data}. Bottom panel: total dust emission at 50 GHz 
  obtained by adding the contribution from the two dust components
smoothed with a 7 degree Gaussian beam (non-linear scale). The CMB and dust
reconstructions have been plotted in the same scale as the
corresponding reconstructions of Fig.~\ref{fig:rec_data}. Units are
$\mu$K (thermodynamical temperature).}  
\end{figure}

We have tested this possibility by adding to our set of data the
second COBE-DIRBE channel (at 2141 GHz) and reconstructing two dust
components. Since we have a total of 6 spectral parameters, it is not
possible to cover exhaustively the full range of possible values.
Instead, we have fixed the values of $\beta_{\rm ff}$,
$\beta_{\rm syn}$ and $\alpha_d$ to the ones found to be optimal in
$\S\ref{sec:freq_data}$ ($-0.19$, $-0.8$ and 2 respectively) 
and varied the dust temperatures of the two dust 
components in the range of 4 to 22 K. The best
reconstruction was found for $T_1=20$ K and $T_2=11$ K with a value of
$G=53.9005$. The CMB recovered map is given in the top panel of
Fig.~\ref{fig:2dust} and has been plotted in the same scale as the one of
Fig.~\ref{fig:rec_data} to allow for a straightforward comparison. The
middle panel shows the difference between the reconstructed CMB map
using one dust component minus the one obtained with two dust
components. It is obvious that the dispersion of the CMB map has been lowered
inside the Galaxy (47.4 $\mu$K in the one dust component versus 42.4
$\mu$K in the two dust components) and the new CMB reconstruction seems
to have less galactic contamination. This can also be seen in
Fig.~\ref{fig:ps_data_cmb} which shows the power spectra of both
reconstructions as well as the COBE-DMR data points. 
The CMB reconstruction obtained
using two dust components is closer to the measured power spectra.
However, both maps are very similar outside the galactic cut. The
dispersion of the difference map in this region is 5 $\mu$K, which is much 
lower than the 
estimated statistical error. The free-free and synchrotron reconstructed maps
are very similar for both models, with differences $\sim$3 per
cent for the free-free and $\sim$2 per cent for the synchrotron.
The dust emission is now composed of the sum of two dust components.
Each of them may not have a physical meaning but the extra degree of
freedom should help to account for temperature variations of the dust
component. In fact, the cold dust component presents negative as well
as positive values in the reconstructed map and should be understood
as a correction to the hot component, which helps to fit both of the
COBE-DIRBE maps at high frequencies as well as some excess emission
at lower frequencies. The total dust emission at 50 GHz is given in
the bottom panel of Fig.~\ref{fig:2dust}. Note that the emission in the Galaxy
centre has been extended with respect to the case when only a single
dust component was used. Basically, the excess of emission of the CMB
reconstructed map in the one single component case is being put into
this cold dust component. Although this seems to improve the results,
the normalised residual noise maps for the COBE-DMR frequency channels present
some structure in the galactic centre and they look very similar to
those obtained for the case of one single dust component (bottom panel
of Fig.~\ref{fig:resnoise02}). Therefore,
our model is still not fitting the Galaxy properly and a more
exhaustive analysis should be performed in order to determine the
correct spectral dependences of the components of the microwave sky. 

Another possible explanation for the difficulty in fitting the data with
our model is the presence of an unknown component. As already
mentioned, several works have found an
anomalous galactic emission at low frequency, which correlates
with the thermal dust one.
Although in principle a free-free component is fitting the
data that we have used, its amplitude is higher 
than expected from estimations based on the H$_\alpha$ emission.
The spectral index found for the free-free was also slightly higher
than expected 
from theoretical predictions. As a test, we have also run the code for
a few models with $\beta_{\rm ff}=-0.22$, two of which gave
slightly lower values of $G$ than the chosen model. However, the
synchrotron map was presenting negative features in the
Galaxy, which is indicative of a degeneracy between the spectral
parameters.

To account for this anomalous component,
we have tried to fit for a spinning
dust component using the extra COBE-DIRBE channel 
in addition to the CMB, free-free, synchrotron and
(thermal) dust emissions. The spectral parameters have been fixed
to the values found in
$\S\ref{sec:freq_data}$. We have tried three different spinning dust
models which correspond to the cold neutral medium, the warm neutral
medium (described in Draine \& Lazarian, 1998) and a combination of
both. In the three considered cases, the $G$ value was appreciably higher
than in the optimal case and the CMB reconstruction was lost.
However, even if spinning dust were present in the data, it
would have been very difficult to recover it. 
We are basically trying to reconstruct three components, CMB, free-free
and spinning dust from just the 3 COBE-DMR channels, which are very
noisy, and therefore the data just do not have enough information.
Moreover, a wider range of frequencies (between 5 and 60 GHz) would be
necessary to distinguish between spinning dust and the other galactic
components.

\medskip
Finally, we would like to point out that 
the combination of MEM and other reconstruction methods
should also be investigated in the future. In particular,
blind source separation methods can infer the spectral dependence of
the reconstructed components under certain assumptions. This
information could be used as input for the MEM algorithm, which would 
improve the results of each method alone.

\section*{Acknowledgements}
RBB thanks the Ministerio de Ciencia y Tecnolog\'\i a and the
Universidad de Cantabria for a Ram\'on y Cajal contract. PV
acknowledges support from a fellowship of Universidad de Cantabria.
We would like to thank G. Giardino for help with foreground
templates and D.P.Finkbeiner, M.Davis and D.J.Schlegel 
for providing the destriped Haslam map.
We thank the RTN of the EU project HPRN-CT-2000-00124 for partial
financial support. RBB and PV acknowledge partial financial support
from the Spanish MCYT project ESP2002-04141-C03-01.
This work has used the software package HEALPix (Hierarchical, Equal
Area and iso-latitude pixelization of the sphere,
http://www.eso.org/science/healpix), developed by K.M. Gorski,
E. F. Hivon, B. D. Wandelt, J. Banday, F. K. Hansen and
M. Barthelmann. The Wisconsin H-Alpha Mapper is funded by the National Science
Foundation.

\appendix
\section{$\chi^2$ function and derivatives}
\label{derivatives}

In order to evaluate the $\chi^2$ in real space we need to be able 
to predict the data at each pixel from the hidden variables ${\mathbf h}\lm$.
The temperature field at a given pixel is normally written in terms of
spherical harmonic coefficients. So, at a position $i$ corresponding to
angles $(\theta,\phi)$ and at
frequency $\nu$, the predicted noiseless data $d^p_{\nu}$ can be written as
\begin{equation}
d^p_\nu(\theta,\phi)=\sum_{\ell=0}^{\ell_{max}} \sum_{m=-\ell}^\ell a\lm^\nu
Y\lm (\theta,\phi),
\end{equation}
where $a\lm^\nu$ are the predicted harmonic coefficients at frequency
$\nu$ and the $Y\lm$ are the spherical harmonics:
\begin{equation}
Y\lm(\theta,\phi)= \left[ \frac{2\ell+1}{4 \pi}
\frac{(\ell-m)!}{(l+m)!}\right]^{1/2} P\lm(\cos \theta) e^{im\phi}.
\end{equation}
Taking into account that, for a real field, we have $a_{\ell-m}=(-1)^m
a\lm^*$, we can write $d_\nu$ as
\begin{eqnarray}
d^p_\nu(\theta,\phi) & = & \sum_{\ell=0}^{\ell_{max}} \sum_{m=0}^\ell
k_m \left[ \Re(a\lm^\nu) \cos(m \phi) \right. \nonumber \\
&& -\Im(a\lm^\nu) \sin(m \phi)
\left. \right] \Lambda\lm
\end{eqnarray}
where $k_m=1$ for $m=0$ and $k_m=2$ otherwise. $\Lambda\lm$ denotes
the factor that multiplies the exponential in the $Y\lm$ expression.
The $a\lm^\nu$ are related to the hidden variables $\mathbf{h\lm}$ by:
\begin{equation}
a\lm^\nu=\sum_\nu B_\ell^\nu ({\mathbf F} {\mathbf L}_\ell {\mathbf h}\lm)_\nu,
\end{equation}
where we have assumed that the beam is spherically
symmetric. $B_\ell^\nu$ is the beam in harmonic space at
frequency $\nu$, ${\mathbf F}$ is the conversion $n_f \times n_c$ matrix. which
encodes the frequency dependence of the components, ${\mathbf L}_\ell$
is the Cholesky
decomposition (an $n_c \times n_c$ matrix) and ${\mathbf h}\lm$ is the
column vector containing the hidden modes for each of the microwave
components. 
The $\chi^2$ function is then written as
\begin{equation}
\chi^2=\sum_{\nu=1}^{n_f} \sum_{i=1}^{N_{p}} \frac{ \left[d_{\nu,i}^o
-d^p_{\nu,i}({\mathbf h}\lm)  \right]^2  }{\sigma_{\nu,i}^2}, 
\end{equation}
where $N_{p}$ is the number of pixels of each data map,
$d_{\nu,i}^o$ is the observed data at frequency $\nu$ and pixel $i$
and $\sigma_{\nu, i}$ is the noise dispersion at frequency $\nu$ and
pixel $i$.

If we denote $r\lm^c$ and $j\lm^c$ as the real and imaginary parts of
the hidden harmonic coefficients for component $c$, we can write the
first derivatives of $\chi^2$ as
\begin{eqnarray}
\frac{\partial\chi^2}{\partial r\lm^c} & = & -2 k_m \sum_{\nu=1}^{n_f}
\sum_{i=1}^{N_{p}} \frac{d_{\nu,i}^o
-d^p_{\nu,i}}{\sigma_{\nu,i}^2} \nonumber \\
&& \times \cos(m \phi_i) \Lambda\lm
B_\ell^\nu (FL)_{\nu c},
\nonumber \\
\frac{\partial\chi^2}{\partial j\lm^c} & = & 2 k_m \sum_{\nu=1}^{n_f}
\sum_{i=1}^{N_{p}} \frac{d_{\nu,i}^o
-d^p_{\nu,i}}{\sigma_{\nu,i}^2} \nonumber \\ && 
\times \sin(m \phi_i) \Lambda\lm
B_\ell^\nu (FL)_{\nu c}.
\end{eqnarray}
The second derivatives are given by
\begin{eqnarray}
\frac{\partial^2\chi^2}{\partial r\lm^{c} \partial r\lm^{c'} } & = 
& 2 k_m^2 \sum_{\nu=1}^{n_f} \sum_{i=1}^{N_{p}} \frac{1}{\sigma_{\nu,i}^2} 
\left[ \cos(m \phi_i) \Lambda\lm B_\ell^\nu \right]^2 \nonumber \\
 && \times (FL)_{\nu c} (FL)_{\nu c'}, \nonumber \\
\frac{\partial^2\chi^2}{\partial j\lm^{c} \partial j\lm^{c'} } & = 
& 2 k_m^2 \sum_{\nu=1}^{n_f} \sum_{i=1}^{N_{p}} \frac{1}{\sigma_{\nu,i}^2} 
\left[ \sin(m \phi_i) \Lambda\lm B_\ell^\nu \right]^2 \nonumber \\
&& \times (FL)_{\nu c} (FL)_{\nu c'}. 
\end{eqnarray}
These second derivatives of $\chi^2$ are the main contribution to the
second derivatives of $\varphi=\chi^2-\alpha S$, since 
the term coming from the entropy is generally
much smaller. They are also the main contribution to the Hessian,
since we are assuming that we can neglect the rest of the elements. 
Note that these derivatives are all positive and
therefore $\varphi$ will have a well defined minimum in the $\mathbf{h\lm}$
space. This means that the Newton-Raphson (NR) 
method will always be driven in the
right direction and should be able to find the desired global minimum.
Note also that the second derivatives of $\chi^2$ depend only on the
characteristics of the experiment, on the assumed spectral
parameters and also in the initial power spectra 
but they are independent of the $h\lm$. Therefore we do not need to
reevaluate these derivatives at each NR iteration.

\end{document}